\documentclass[a4paper,usegraphicx,usenatbib]{mn2e}

\usepackage{times}
\usepackage{amsmath}
\usepackage{amssymb}
\usepackage{epsfig}

\begin{document}

\title[Black Hole and Stellar Population of MCG--6-30-15]{The Black Hole and Central Stellar Population of MCG--6-30-15}
\author[S.I. Raimundo et al.]
{\parbox[]{7.in}{S.~I. Raimundo$^{1}$\thanks{E-mail: 
sandra.raimundo@sissa.it}\footnotemark, R.~I. Davies$^{2}$, P. Gandhi$^{3}$, A.~C. Fabian$^{4}$, R.~E.~A Canning$^{4}$, V.~D. Ivanov$^{5}$\\[0.2cm]
\footnotesize
$^{1}$SISSA - International School for Advanced Studies, via Bonomea, 265, 34136 Trieste, Italy\\
$^{2}$Max-Planck-Institut f\"{u}r extraterrestrische Physik, 85741 Garching, Germany\\
$^{3}$Institute of Space and Astronautical Science (ISAS), JAXA, 3-1-1 Yoshinodai, chuo-ku, Sagamihara, Kanagawa 229-8510, Japan\\
$^{4}$Institute of Astronomy, Madingley Road, Cambridge CB3 0HA\\
$^{5}$European Southern Observatory, Ave. Alonso de Cordova 3107, Vitacura, Santiago 19001, Chile}
}

\maketitle
\begin{abstract}
We present the first near-infrared integral field spectroscopy observations of the galaxy MCG--6-30-15. The \emph{H}-band data studied in this paper cover the central 500\,pc of the galaxy at the best resolution (0${''}$.1) so far. The spectra of the innermost regions are dominated by broad brackett series emission lines and non-stellar continuum, under which we are able to trace the distribution and kinematics of the stars and also the [Fe\,II] line emission.
We find that there is a counter-rotating stellar core extending out to 125\,pc, which appears to be associated with the [Fe\,II] emission. Based on the mass-to-light ratio, and the presence of this emission line, we estimate the age of the central stellar population to be of order of 65\,Myr. We show that the gas needed to fuel the black hole is, at most, only 1 per cent of that needed to form these stars.
We derive independent constraints on the black hole mass using the dynamical information and determine an upper limit for the black hole mass, $M_{\rm BH} < 6\times 10^{7}$M$_{\odot}$, that is consistent with other estimates.
\end{abstract} 
\footnotetext{Based on observations collected at the European Organisation for Astronomical Research in the Southern Hemisphere, Chile, during program 077.B-0553(B).}
\section{Introduction}
The galaxy MCG--6-30-15 is an elongated lenticular (S0) galaxy and classified as a Seyfert 1.2 \citep{boisson02}. It presents a dust lane south of the nucleus and parallel to the photometric major axis of the galaxy \citep{malkan98,ferruit00}. The spectral energy distribution (SED) peaks in the mid-IR and is probably due to thermal emission from warm/hot dust grains.
In the optical, the spectrum is dominated by broad Balmer lines and narrow forbidden oxygen lines (including [O III] $\lambda5007$ \AA), and the narrow line region appears to have a large spatial extent \citep{bennert06}. 
\begin{figure*}
\begin {centering}
\includegraphics[width=1.7\columnwidth]{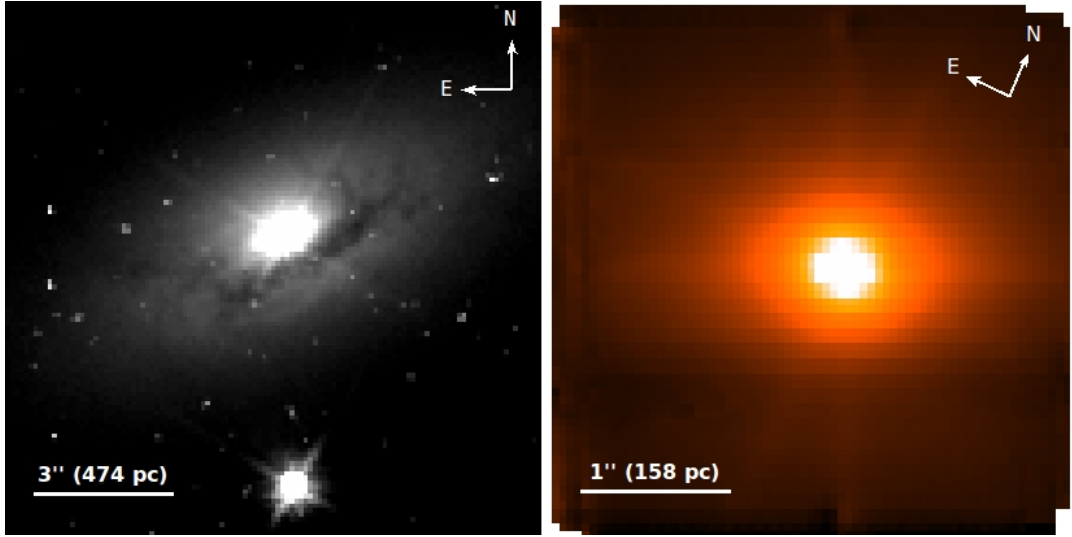}
\caption {HST and SINFONI images of MCG--6-30-15. Left: WFC2 broadband image of MCG--6-30-15 from \citealt{malkan98}. Right: SINFONI \emph{H}-band data cube collapsed to two dimensions for plotting purposes. The SINFONI field-of-view is rotated by 25 degrees clockwise relative to the HST image.}
\label{hst_sinfoni}
\end{centering}
\end{figure*}
Strong internal reddening is observed ($E(B-V) \sim (0.5 - 1.0)$ \citealt{reynolds97,boisson02}) and the spectrum show signs of complex obscuration (\citealt{reynolds97,ballantyne03}). \cite{reynolds97} do a multi-wavelength analysis of this galaxy, with optical, UV, IR and X-ray data and find evidence of a dusty warm absorber (i.e. a column of dusty ionised material, usually in the form of an outflowing wind) along our line-of-sight. 
\cite{ballantyne03} argue that the dust lane could be responsible for the reddening observed and for part of the warm absorber.
MCG--6-30-15 has an Active Galactic Nucleus (AGN) which has been studied extensively in the X-ray band (L$_{\rm X} (2-10 \rm keV)$ $\sim4\times10^{42}$ erg s$^{-1}$ \citealt{winter09, vasudevan09}). The non-stellar continuum can also be observed in the infrared \citep{oliva99} and is detected but only marginally resolved in the radio \citep{nagar99,mundell09}. High angular resolution observations in the mid-IR \citep{horst09, gandhi09}, reaching resolutions of the order of $\sim 0{''}.35$, show the nucleus as a point source.

This galaxy is best-known for having provided the first detection of a relativistic broadened Fe K$_{\alpha}$ emission line generated in the inner regions of an accretion disc \citep{tanaka95}, hence providing evidence for the presence of a supermassive black hole. The X-ray spectral variability has been studied in detail (e.g. \citealt{reynolds95, nowak&chiang00, uttley02, vaughan&fabian04, mchardy05, emmanoulopoulos11}) and has provided approximate constraints for the mass of the black hole, assuming that the break frequency scales with the black hole mass. The black hole spin has been measured from the spectral fitting of the Fe K$_{\alpha}$ emission line to be $a = 0.989$ \citep{brenneman06}, i.e. a rapidly rotating black hole.

Despite the detailed studies mentioned above, the mass of the black hole responsible for the AGN activity is still not very well constrained. The galaxy is too distant for current instruments to resolve the black hole sphere of influence. Recent work to measure the black hole mass of MCG--6-30-15 has been done by \cite{mchardy05} who discuss several methods (M$_{\rm BH}-\sigma$ relation, virial type relations and spectral variability) to determine the mass.
They measure the stellar velocity dispersion from the Ca II absorption lines up to a radius of $R_{e}/8$ and obtain a value of $\sigma = 93.5$\thinspace km\thinspace\thinspace s$^{-1}$. Using the various $M_{\rm BH} - \sigma$ relations available \citep{tremaine02,ferrarese02} and constraints from variability, the mass is found to be in the range $M_{\rm BH} \sim (3 - 6) \times 10^{6}$ M$_{\odot}$. Another study \citep{bennert06}, finds somewhat larger black hole masses $(0.8 - 3) \times 10^{7}$ M$_{\odot}$, after using several methods to determine it (virial determinations and M$_{\rm BH}$-$\sigma$ relation). More recently, \cite{winter09} used the method based on the M$_{\rm BH} - $L$_{\rm K}$ relation outlined in \cite{mushotzky08} to calculate a black hole mass of M$_{\rm BH} = 2.3\times 10^{7}$M$_{\odot}$. \cite{vasudevan09} refined this method and determined a similar black hole mass: M$_{\rm BH} = 1.8\times 10^{7}$M$_{\odot}$.

In terms of the stellar population, \cite{boisson04} finds the stellar content to be dominated by an old population. From analysis in the ultra-violet, \cite{bonatto00} find a dominant old bulge stellar population ($\sim 10$ Gyr) with indications of a series of previous bursts of star formation distributed in age among the younger stellar population ($<500$ Myr). Optical analysis by \cite{bennert06} reach the same conclusion.

The advent of integral field spectroscopy (IFS) has opened a new perspective on the black hole fuelling process and on the relation between the stellar properties of the galaxies and the AGN activity.
Studies using this technique have unveiled features in the gas kinematics which are likely due to the dynamics of inflowing and outflowing gas. 
Observations of local Seyfert galaxies show evidence for gas inflowing or outflowing on a scale of tens to hundreds of parsecs (e.g.: \citealt{prieto05}, \citealt{fathi06}, \citealt{storchi-bergmann07}, \citealt{muller-sanchez11}). 
Combined with adaptive optics, IFS provides the opportunity to map with high resolution the inner regions of nearby galaxies and determine the mass of the central black hole (e.g. NGC3227 - \citealt{davies06}, Centaurus A - \citealt{neumayer07, cappellari09}, Fornax A - \citealt{nowak08}, NGC524 and NGC2549 - \citealt{krajnovic09}, M87 - \citealt{gebhardt11}). We can also employ IFS to investigate the AGN nature (e.g. \citealt{valencia-s12}) and the relationship between the star formation and the AGN activity (e.g. \citealt{daviesetal07}). The high level of detail and amount of information that comes out of these observations makes them a very useful probe of the physics in the central regions of galaxies.

MCG--6-30-15 is an ideal system in which to investigate the physics of black hole accretion, but there are still many aspects about its nature that remain unknown. By observing the inner regions of this galaxy, we can study its stellar and dynamical properties and draw conclusions on the black hole and AGN activity.
In this work we present the first study of this galaxy using IFS data in the \emph{H}-band, and determine the kinematic properties in the central regions of MCG--6-30-15. Section \ref{sec:analysis} gives the details on the data reduction. In Section \ref{sec:discussion} we present the results obtained for the stellar kinematics and gas dynamics traced by the [Fe II] emission line. We use dynamical arguments to constrain the black hole mass and the star formation history of the galaxy. The results are summarised in Section \ref{sec:conclusions}.

We adopt the standard cosmological parameters of H$_0 = 70$ km s$^{-1}$ Mpc$^{-1}$, $\Omega_{\rm m} = 0.27$ and $\Omega_{\rm \Lambda} = 0.73$. The scale for our measurements is approximately 0.158 kpc/'' at a redshift $z = 0.0077$ (\citealt{fisher95} value quoted at the NASA/IPAC Extragalactic database). The distance to MCG--6-30-15 used in this paper is 33.2 Mpc \citep{wright06}.\\
\section{Data Analysis}
\label{sec:analysis}
\subsection{Data reduction}
\begin{figure}
\begin {centering}
\includegraphics[width=1.0\columnwidth]{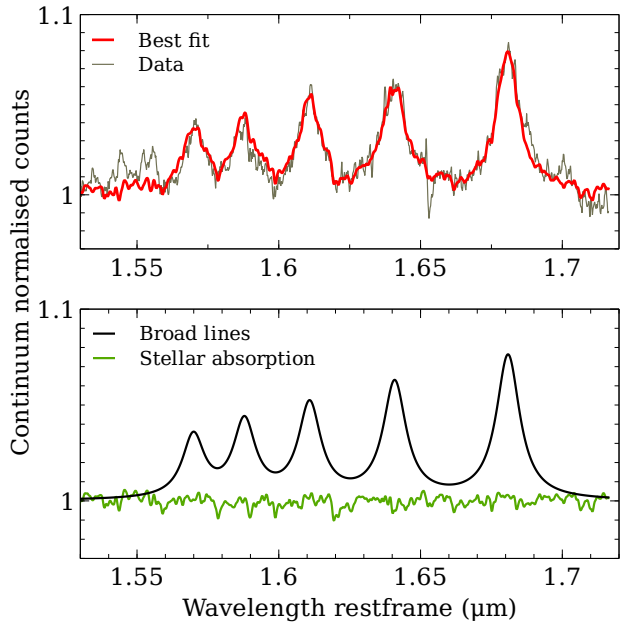}
\caption [Decomposition_spectra] {Spectral data integrated in the central 6 x 4 pixel region of the field-of-view and the best fit spectral components as a function of the rest-frame wavelength. Top: Data (black) and best fit model (red). Bottom: Spectral components used in the fit, broad Brackett emission lines (black) and fitted sum of stellar templates tracing the absorption features (green). The scale is normalised to the continuum value. The excess at around 1.55 $\micron$ corresponds to the position of another broad line in the Brackett series. We tried to extend the fit to this line with a sixth Brackett emission line but due to its low intensity the fit was not very good. We decided then not to fit this line and exclude its wavelength range from the subsequent data analysis.}
\label{decomposition_spectra}
\end{centering}
\end{figure}
We obtained spectroscopy data for our source, with the near-IR integral field spectrograph SINFONI on the VLT (\citealt{eisenhauer03,bonnet04}), using a natural guide star for the adaptive optics (AO). The observations were carried out in the \emph{H}-band mainly to avoid the contamination from the non-stellar continuum. In the near-infrared, the relative contribution of the non-stellar emission compared with the stellar emission is lower than in other wavelengths and less affected by extinction. The \emph{H}-band, in particular, shows less nuclear emission reprocessed by dust than in the \emph{K}-band \citep{boisson02}. The \emph{H}-band spectra with R$\sim$3000, were taken on three nights, (April 2006, 4th, 5th and 20th), with a pixel scale 0${''}$.05''$\times$0${''}$.1, sampling a field-of-view of 3''$\times$3'' (right panel of Fig.~\ref{hst_sinfoni}). Each exposure was of 300 seconds, alternating between the object (A) and the sky (B) in the sequence ABBA. This sequence was repeated eight times. The total on-source exposure time is 1h20m.

The data were reduced using the ESO pipeline for SINFONI 
version 3.8.3, and independent routines when necessary. 
Using the pipeline and the calibration files, we obtained the bad pixel maps (non-linear pixels, hot pixels and reference bad pixels), master dark and master flat frames (recipes {\sc sinfo\_rec\_detlin}, {\sc sinfo\_rec\_mdark} and {\sc sinfo\_rec\_mflat}). These routines fail to identify some of the bad pixels, so we visually inspected our raw frames and manually selected extra bad pixels, in particular horizontal bad pixel lines that can be seen in the first pointings of each of the four ABBAABBA sets of observations.

We subtract the master dark from the raw object and sky frames, and use the recipe \textsc{sinfo\_rec\_jitter} to correct for bad-pixels and distortions, flat-field the data, wavelength calibrate and reconstruct 32 data cubes for object and sky. The \textsc{sinfo\_rec\_jitter} routine is run with the sky subtraction options turned off, to allow us to correct the object data cubes for sky emission following the method described in \cite{davies07}. From this routine we obtain each input data cube corrected for the sky emission.
Our final 16 object data cubes are then checked for any remaining bad pixels using a 3D Laplacian edge detection IDL code \citep{davies10}, 
which is a 3D version of LA cosmic \citep{vandokkum01}.

In the near-IR, the atmosphere greatly affects the transmission and there are several telluric absorption signatures visible in the spectrum. We observed four telluric standard stars in the \emph{H}-band but we found those were not suitable due to low signal-to-noise ratios. Instead, we divide our data cubes by a normalised theoretical template telluric spectrum re-scaled to the air-mass of our observations.
The last step was to combine the 16 individual exposure cubes into a final cube. We re-sample each cube to a half-pixel scale both in $x$ and $y$ dimensions and offset them using the values recorded in their headers (\textsc{cumoffset} and \textsc{tel targ alpha/delta}), verifying that the bright galactic nucleus is aligned in every spatially shifted cube, and then median combine them. The final data cube (collapsed to two dimensions for plotting purposes) is shown in the right panel of Fig.~\ref{hst_sinfoni}. The pixel scale of the final reduced cube is 0${''}$.05 $\times$ 0${''}$.05 and the horizontal $x$-axis is aligned with the galaxy's major axis, which means a tilt of $\sim25^{\circ}$ in relation to the East-West direction.
\subsection{Flux calibration}
\begin{figure}
\begin {centering}
\includegraphics[width=1.0\columnwidth]{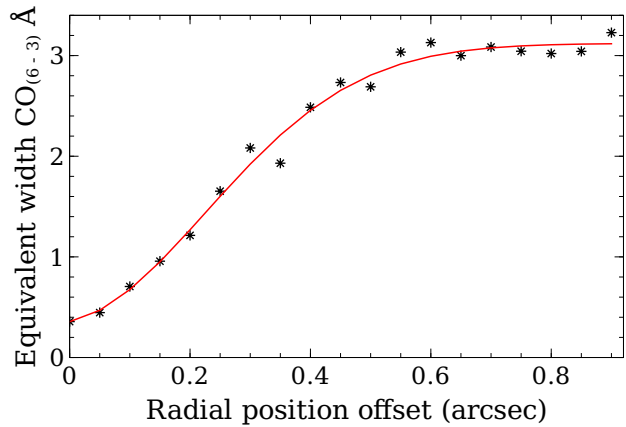}
\caption [Eq width] {Equivalent width of the CO (6-3) 1.6187 $\micron$ absorption line measured in integrated annuli at distance $r$ from the AGN position. The change in equivalent width is due to non-stellar continuum associated with the AGN.}
\label{eqwidth}
\end{centering}
\end{figure}
\begin{figure*}
\begin {centering}
\includegraphics[width=2.0\columnwidth]{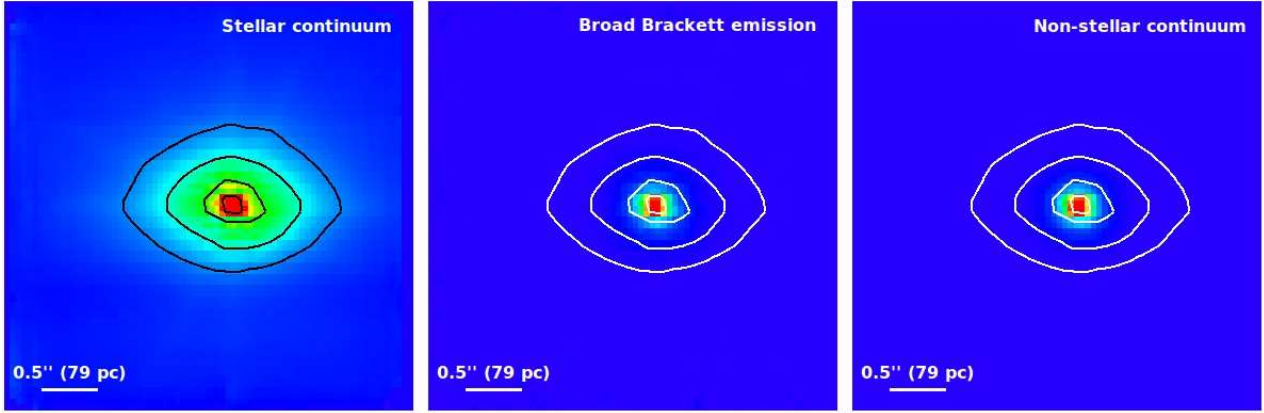}
\caption [Decomposition] {Panel with decomposed emission. Left: Stellar continuum. Centre: Broad Hydrogen Brackett emission lines. Right: Non-stellar continuum. Stellar continuum contours are overlapped on the three panels and provide a comparison between the stellar distribution and the unresolved AGN components (broad Brackett emission lines and non-stellar continuum).}
\label{decomposition}
\end{centering}
\end{figure*}
The flux calibration of our data cube is done based on the 2MASS \emph{H}-band image of the galaxy, since our standard stars were not suitable (low S/N or saturated).
We integrate the counts in a 3'' diameter circular aperture in the 2MASS image, which gives an observed magnitude of m$_H$ = 11.3 and a flux of $3.37\times 10^{-14}$W m$^{-2} \micron^{-1}$ \citep{cohen03}. We then integrate the spectra in our data cube in a 3'' aperture centred at the same physical position as for the 2MASS image. We compare the median count number of our integrated spectrum in the range $\lambda = [1.509 - 1.799]\micron$ with the flux from the 2MASS image and use this as our count to flux calibration. Our calibration was checked using different apertures in 2MASS and is consistent to $\sim 20$ per cent.

\subsection{Emission components}
In the case of MCG--6-30-15, the broad hydrogen Brackett AGN emission lines, $v_{\rm FWHM}\sim 1800$ km/s, are very strong in the central pixels and dominate over the wavelength range of interest, making the stellar absorption features hard to identify. The following method is adopted to remove the broad emission lines. We first select an integrated central 6 by 4 pixel region where the broad lines are stronger and fit the five strongest emission lines with Lorentzian profiles, since these provide a good fit to the lines \citep{veron-cetty01}. We fix the atomic parameters and the velocity width of the line (relative half-width at half-maximum and relative intensity - using as an approximation \citealt{hummer&storey87} for case B recombination). To improve the fitting we also include a second order polynomial function to fit the non-stellar continuum and a giant-star template spectrum convolved with a Gaussian as a basic model to the stellar continuum and its absorption features. The stellar template is of a M1 III star observed previously with SINFONI in the 100 mas scale. It was chosen due to its deep CO absorption lines which are also observed in our spectra. The plot in Fig.~\ref{decomposition_spectra} shows the spectral components fitted to the data. The best-fit parameters for the Brackett lines obtained from the integrated spectrum are used as initial guesses when fitting every spaxel. The spaxel-by-spaxel fitting determines the line parameters for the Brackett emission at each spatial position and removes it from the data cube.

\begin{figure}
\begin {centering}
\includegraphics[width=1.0\columnwidth]{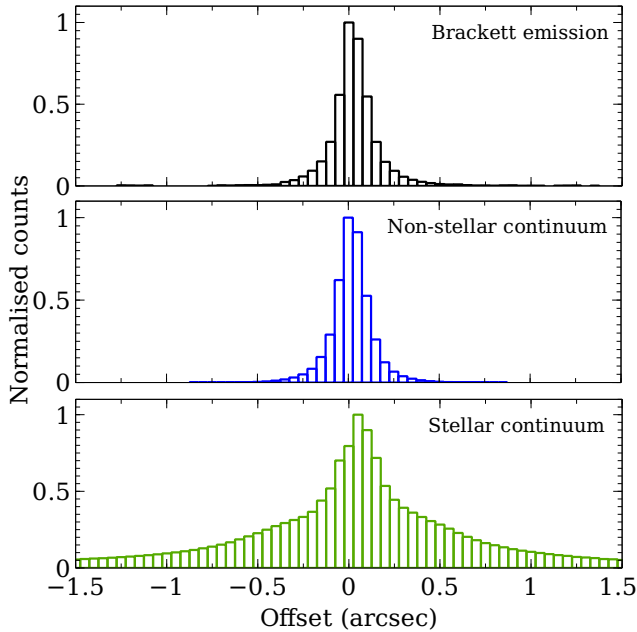}
\caption [Cuts] {Distribution along the major axis of the galaxy. The zero offset corresponds to the AGN position. Top: Brackett emission. Middle: Non-stellar continuum. Bottom: Stellar continuum. The stellar contribution has a broader core and clearly extends to a larger radii than the one from the AGN. The orientation is as for Fig.~\ref{decomposition}.}
\label{cuts}
\end{centering}
\end{figure}

With this approach we can decompose the emission into the individual AGN and stellar contributions. The broad Brackett emission is due to the AGN, while the absorption lines are due to the stellar population. However, the continuum is composed of stellar emission and AGN heated dust. To separate these two contributions we use the equivalent width of one of the strongest absorption lines present in our data: CO (6-3) $\lambda = 1.6187$. The argument is that the deviations of the equivalent width from the intrinsic value will be caused by dilution due to the non-stellar continuum. By measuring the equivalent width at different radial positions and comparing the measurements to an intrinsic value, it is possible to determine the relative AGN and stellar contributions at each position. The stellar fraction will be,
\begin{eqnarray}
f_{\rm stellar}(r)=\frac{W_{\rm obs}}{W_{\rm intr}}(r),
\end{eqnarray}
where $W_{\rm obs}$ is the observed equivalent width and $W_{\rm intr}$ is the intrinsic equivalent width. We measure the equivalent width of the line in integrated radial bins, and plot it as a function of distance from the nucleus in Fig. \ref{eqwidth}. We fit a Gaussian to the distribution and use it to decompose the continuum. The maximum equivalent width ($\sim 3.1 \AA$ at r$ = 0{''}.8$) is taken as the intrinsic value for our case. Although it is lower than what has been observed before ($\sim 4 \AA$) \cite{daviesetal07}, \cite{valencia-s12}, this could be due to the presence of dust farther out in the galaxy. Lower values have also been observed by \cite{oliva99} using slit-spectroscopy on MCG--6-30-15. 
In Fig.~\ref{decomposition} we show the decomposition maps of the observed emission into the separate contributions from the stellar continuum (left), the Brackett broad emission (centre) and the non-stellar continuum (right). In Fig.~\ref{cuts} we show the cuts along the major axis, passing through the AGN position determined by the peak of the Brackett emission (offset = 0 arcsec). From the plot, it is clear that the Brackett emission and the AGN continuum have approximately the same distribution, which is what we would expect since they are both spatially unresolved (\citealt{bennert06} estimate the broad line region to be 9\,light-days across). The stellar continuum shows more extended wings which correspond to the stellar population spatially distributed in the galaxy.

\subsection{PSF and spectral resolution}
We determine the point spread function (PSF) from the Brackett emission map. The broad Brackett emission is coming from a very small region around the AGN and it is not resolved in our data. The spatial extent of the observed emission will be caused by the instrument and the atmosphere and we can measure the PSF by modelling its distribution. In our data the PSF is well modelled by a double Gaussian: a Gaussian with small width to model the peak (FWHM = 1.4 pixels $\sim0{''}.07$) and a wider one with (FWHM = 4.0 pixels $\sim0{''}.2$) to model the wings. The non-stellar continuum can also be used to determine the PSF since the dust-emitting region is unresolved. We obtain similar results: FWHM = 1.4 pixels $\sim0{''}.07$ and FWHM = 3.7 pixels $\sim0{''}.18$.

The spectral resolution is obtained from the sky lines. These emission lines are present in our sky exposures, and their width is due to SINFONI's instrumental broadening. We use the combined sky cube and fit an unblended sky line with a Gaussian. The half width at half maximum (HWHM) is taken as the instrumental broadening ($\sigma_{\rm inst} =$ HWHM / $\sqrt{2\ln(2)}$).
We repeat this procedure for all the spatial positions in a masked cube, to exclude the noisy outer regions of the field-of-view. The mean value for the instrumental broadening is $\sigma_{\rm inst} =60$ km/s (3.2 $A$ at $\lambda\sim1.6187\micron$), with variations between 55 km/s and 67 km/s. The unblended line is selected in a wavelength region around the central wavelength of [Fe II]. However, in the wavelength range of our analysis, we note that the maximum instrumental broadening variation is 5 km/s. 
 
\subsection{Extracting stellar kinematics}
\label{sec:extracting}
The velocity map and stellar velocity dispersion distribution are determined from our fully reduced and broad line subtracted data cube, by fitting the spectra with pPXF \citep{cappellari04,vandermarel&franx93}. This {\sc idl} routine uses a set of stellar templates to find the best weighted spectral combination and line-of-sight velocity distribution to fit the input spectrum. The line-of-sight velocity distribution is based on the form of a Gauss-Hermite series expansion, with the first two moments being the velocity offset and the velocity dispersion. It is also possible to fit higher moments such as the asymmetric and symmetric deviations from a Gaussian (namely the $h3$ and $h4$ parameters which are related to the skewness and kurtosis of the distribution).
In this work we fit only the first two moments of the distribution due to the limited signal-to-noise of our data. 

In the \emph{H}-band, there are not many medium or high resolution stellar templates. Our data have a resolution $R\sim3000$, which is higher than most of the templates found in the literature. For this reason we use different sets of stars to determine the stellar population and the line-of-sight distribution.
The stellar templates of \cite{meyer98}, which include several spectral types and luminosity classes, have spectral resolution of the order of our data resolution ($R\sim3000$), but we encountered problems when determining the velocity properties. We use this library to determine that our spectra can be well fit based on a set of K and M stars.
To determine the velocity properties we use the more recent stellar templates at higher resolution of $R\sim5000$ obtained by \cite{le11}. This library contains only ten G, K and M giant stars, but as we found previously, K and M stars reproduce our spectra well and are suitable to model the stellar absorption features.
To remove the effects of differential instrumental broadening between the SINFONI data and the templates, we broaden the stellar templates (FWHM$_{\rm temp}$ = 8 pixels $\sim 3.2\times10^{-4}\thinspace\thinspace  \micron$) to the resolution of SINFONI by convolving them with a Gaussian kernel ({\sc psf\_gaussian} in {\sc idl}) of $\sigma_{\rm convol} = \sqrt{\sigma^{2}_{\rm inst}-\sigma^{2}_{\rm temp}}$. $\sigma^{2}_{\rm inst}$ is the SINFONI instrumental resolution which we measure from the sky emission lines.
\begin{figure*}
\begin{centering}
\includegraphics[width=1.8\columnwidth]{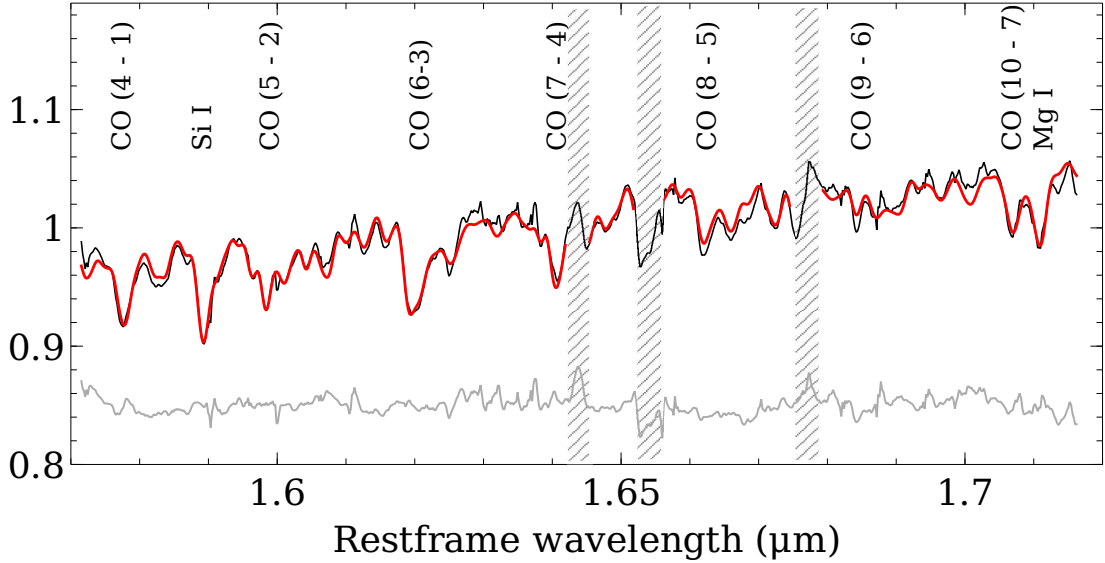}
\caption [Spectrum of central regions and best-fit model]{Spectrum and fitting results versus rest-frame wavelength in $\micron$. In black, integrated spectrum over a radius of $\sim 1{''}.1$ excluding the central 6$\times$4 pixels. In red, pPXF best fit combination of stellar templates and 4th order polynomial (to model the AGN continuum). The residuals are shown in grey and have been offset from zero for plotting purposes. The vertical shaded regions indicate wavebands which were excluded from the fit, including the [Fe II] emission lines at $\lambda = 1.644 \micron$ and $\lambda = 1.677 \micron$ and the absorption feature at $\lambda \sim 1.654 \micron$ (see text for details).}
\label{fit}
\end{centering} 
\end{figure*}
\subsubsection{Binning and masking}
We use the strength of the absorption lines to bin the data spatially, and in this way avoid being dominated by the noise from the AGN continuum. Usually the signal-to-noise is quoted as the 1$\sigma$ noise value in relation to the stellar continuum or total continuum. Here we define our signal-to-noise ratio as the depth of the absorption lines divided by the 1$\sigma$ noise level, since this is ultimately the measurement we are interested in. 
We bin our data using the 2D Voronoi binning routine of \cite{cappellari&copin03} to an average signal-to-noise S/N = 5 (between the amplitude of the CO (6-3) 1.6187 $\micron$ absorption line and the noise RMS in a line free region of the continuum). This corresponds to a S/N $\sim 50$ when measured in relation to the stellar continuum level. A more accurate signal-to-noise measurement will be determined in the next sections from the residuals of our fit.

Individual pixels in the regions where the flux decreases to 1/125 of its peak value were masked out and excluded from the analysis. These pixels in general correspond to the outer regions of our field-of-view. We also masked out a $\sim 6$ pixel wide region parallel to the major axis at around 0${''}$.8 from the centre of the field-of-view, which corresponds to an illumination artefact in SINFONI. 

\section{Discussion}
\label{sec:discussion}
\subsection{Stellar kinematics}
\label{sec:stellarkin}
The stellar kinematic properties are determined by running pPXF on our binned data cube. The input stellar library is the one mentioned above \citep{le11}, and we include an additive polynomial of fourth order to model the AGN continuum. The emission lines of [Fe II] and a residual from the telluric subtraction are masked out when fitting the spectra. The code output includes the best fit spectrum at each pixel, the velocity offset and the velocity dispersion. We do not fit the higher moments $h3$ and $h4$ as mentioned in the previous section due to our low signal-to-noise.
\begin{figure*}
\centering
\epsfig{file=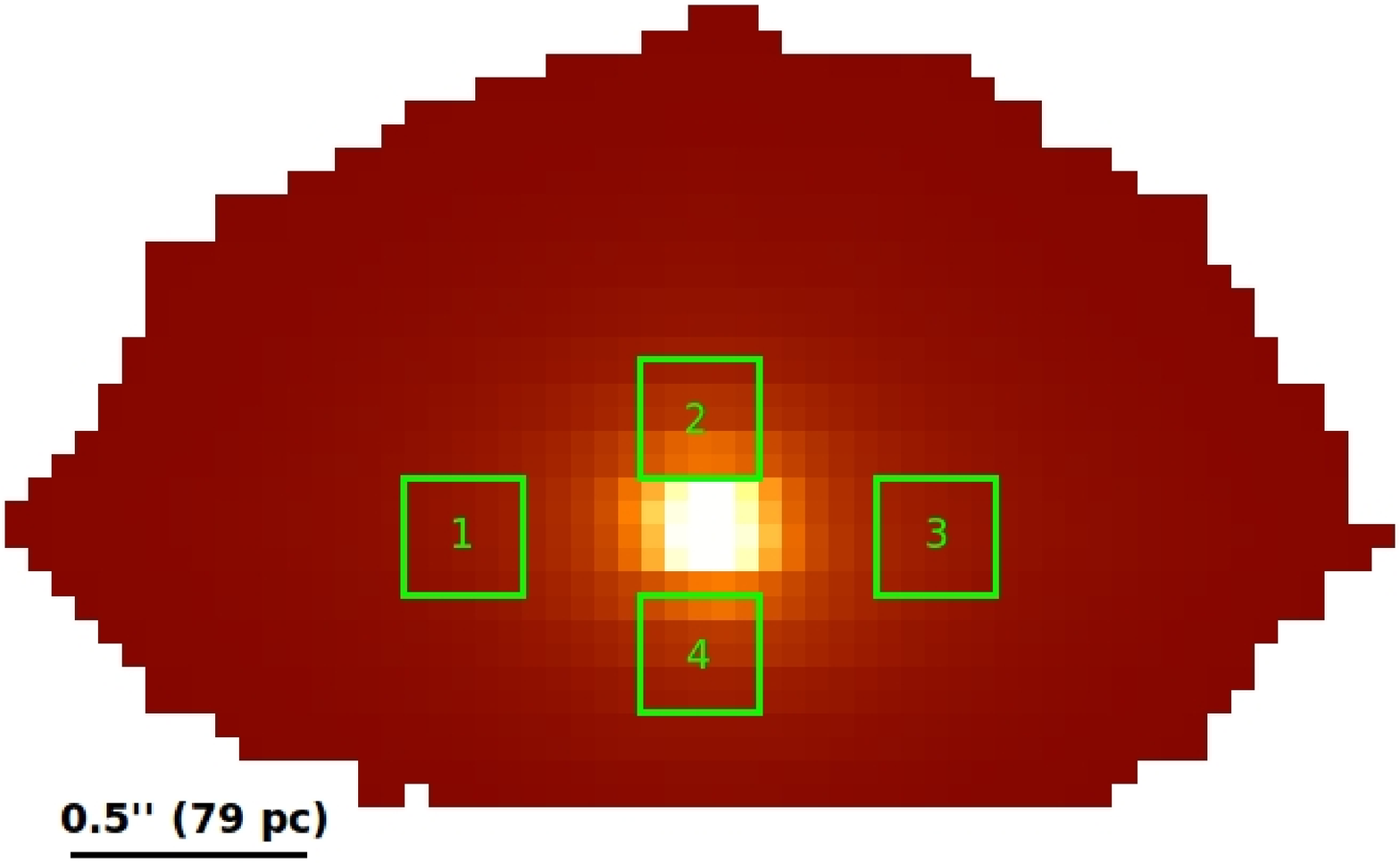,width=0.5\linewidth,clip=}\\[0.5cm]
\begin{tabular}{cc}
\epsfig{file=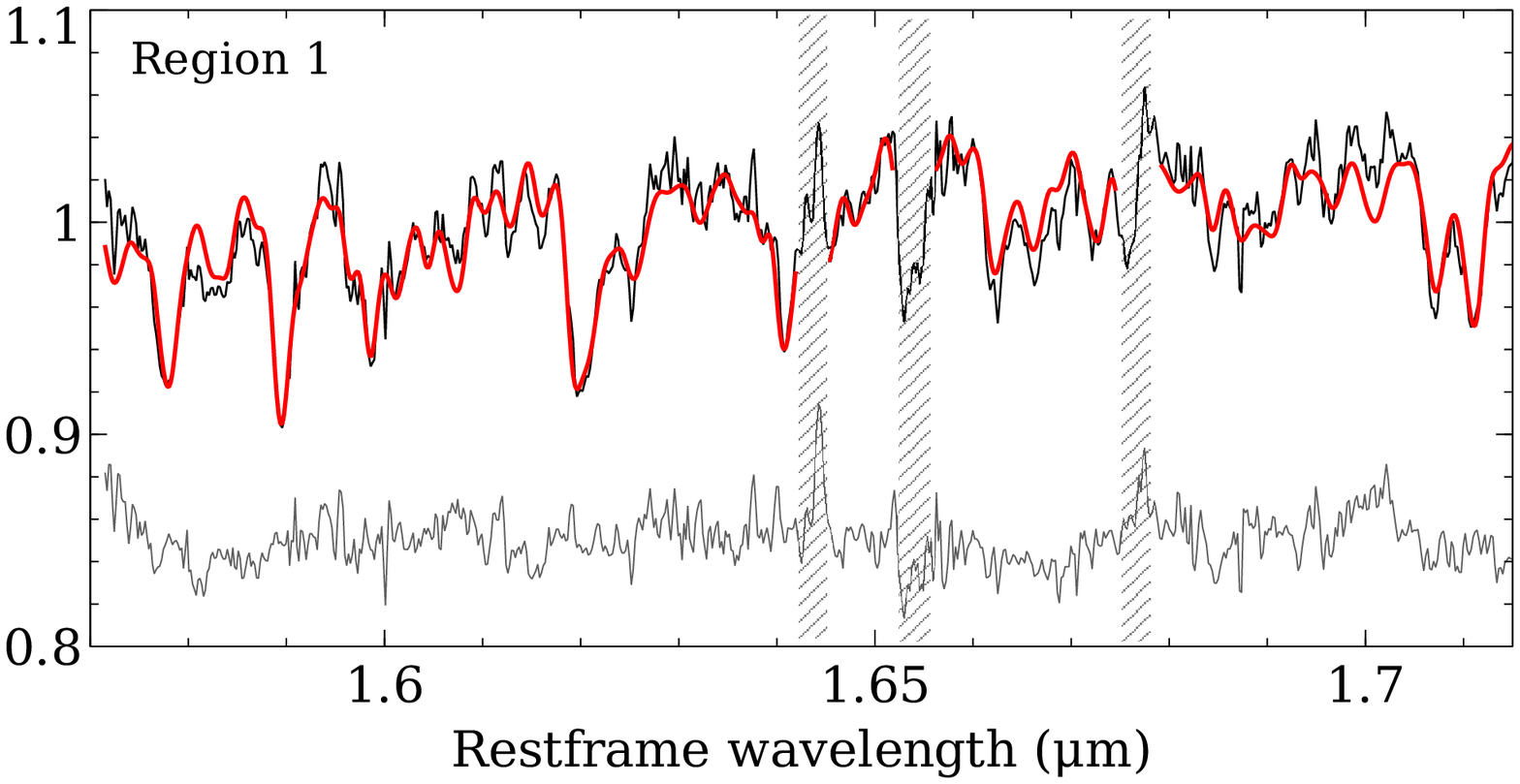,width=0.5\linewidth,clip=} &
\epsfig{file=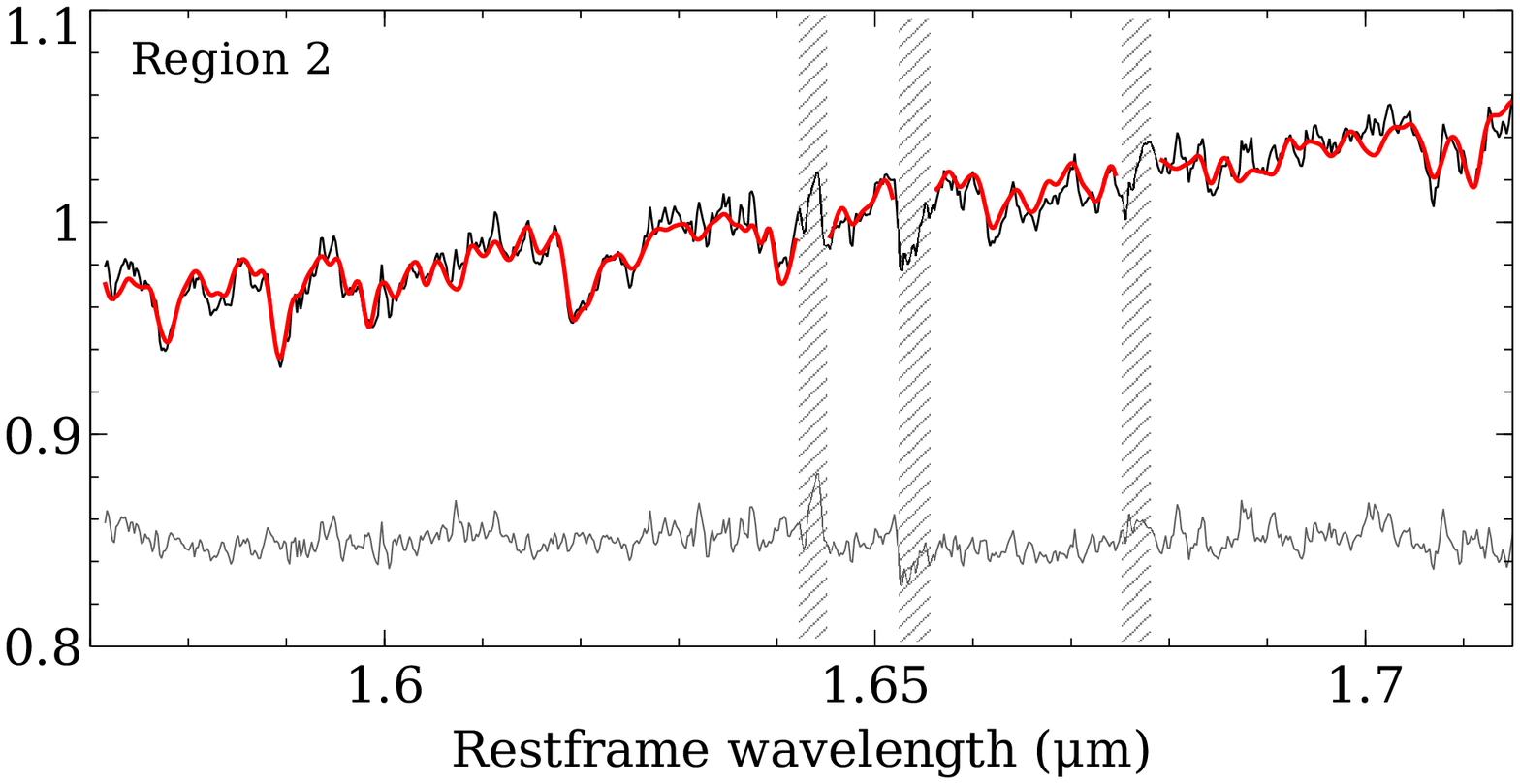,width=0.5\linewidth,clip=} \\[0.5cm]
\epsfig{file=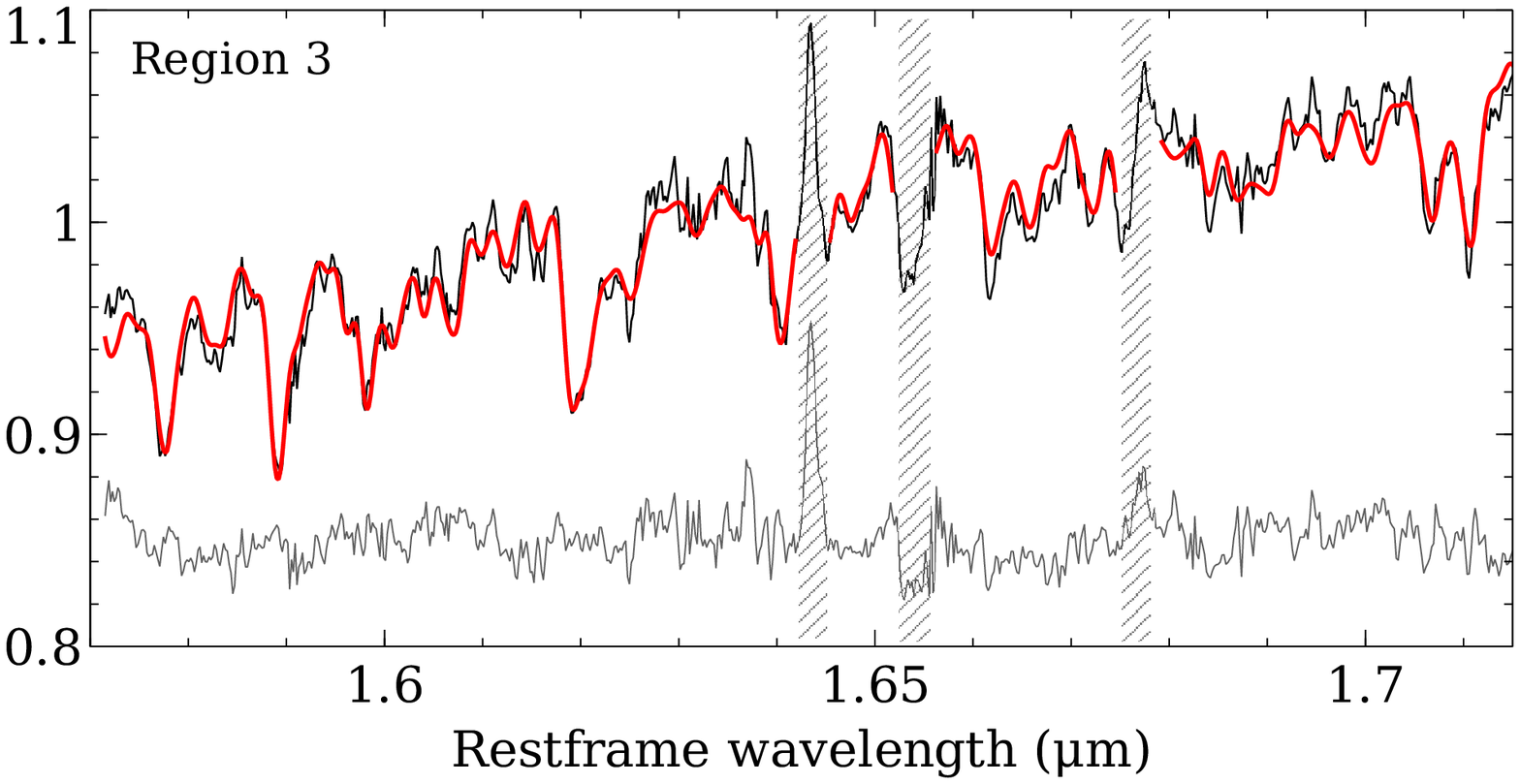,width=0.5\linewidth,clip=} &
\epsfig{file=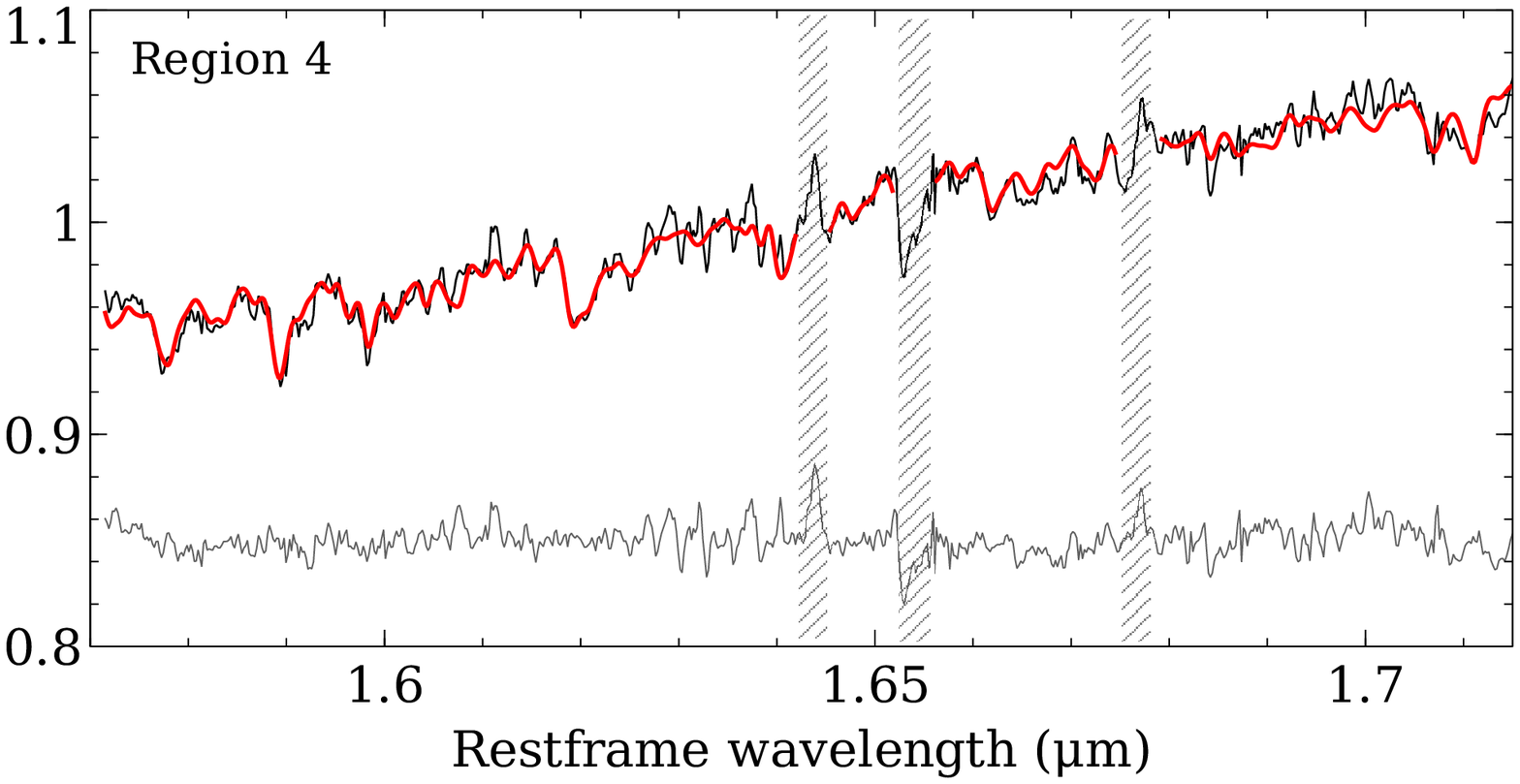,width=0.5\linewidth,clip=}
\end{tabular}
\caption[Regions]{Integrated spectra and best fit model for four different regions of the field-of-view. The size of each region is 5$\times$5 pixels indicated with the green boxes in the top image. The bottom four panels show the spectra (black), the best fit model using pPXF (red) and the residuals (grey) offset for plotting purposes. The vertical shaded bars indicate regions excluded from the fit.}
\label{region_spectra}
\end{figure*}
\begin{figure*}
\centering
\epsfig{file=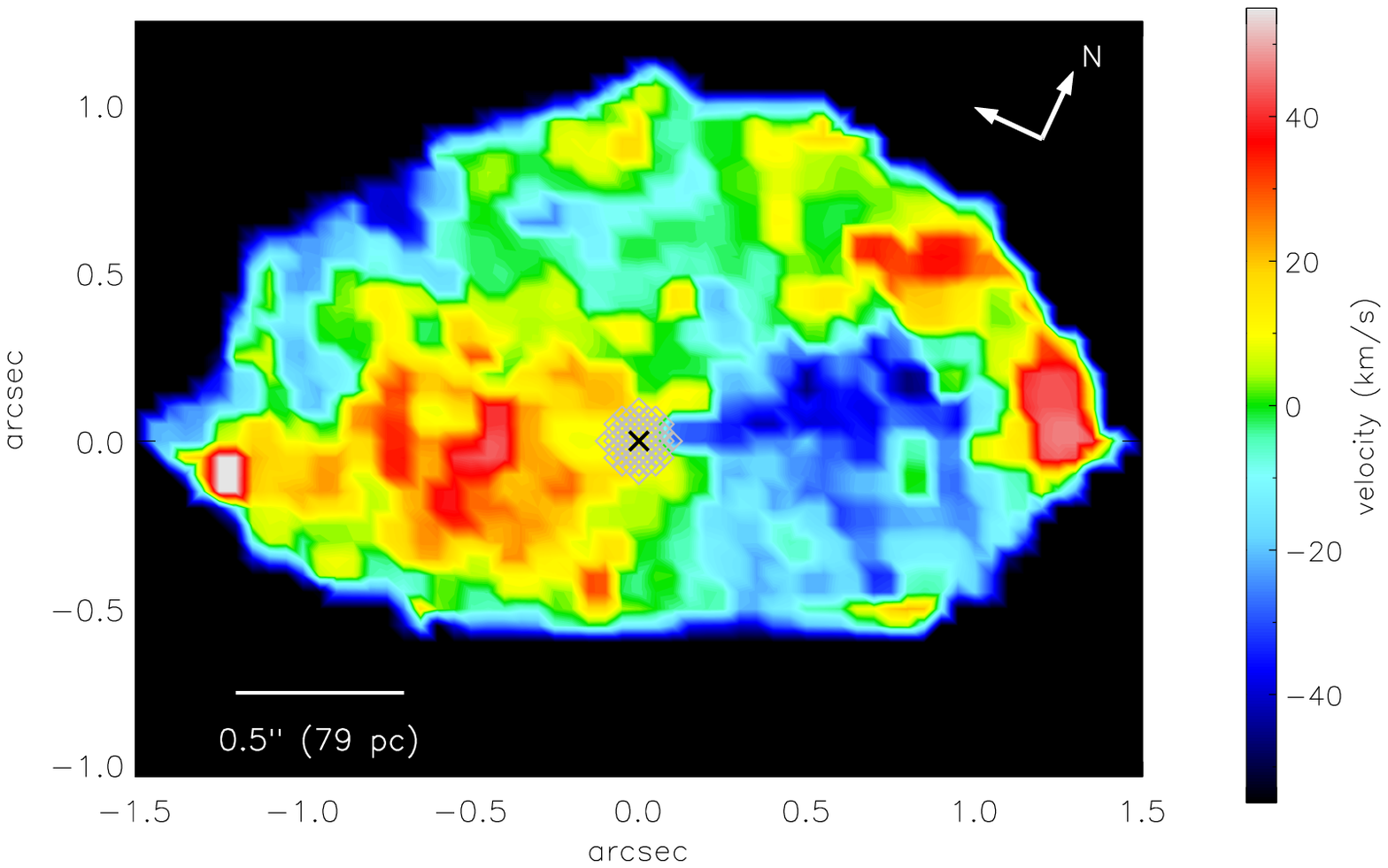,width=0.6\linewidth,clip=}\\
\epsfig{file=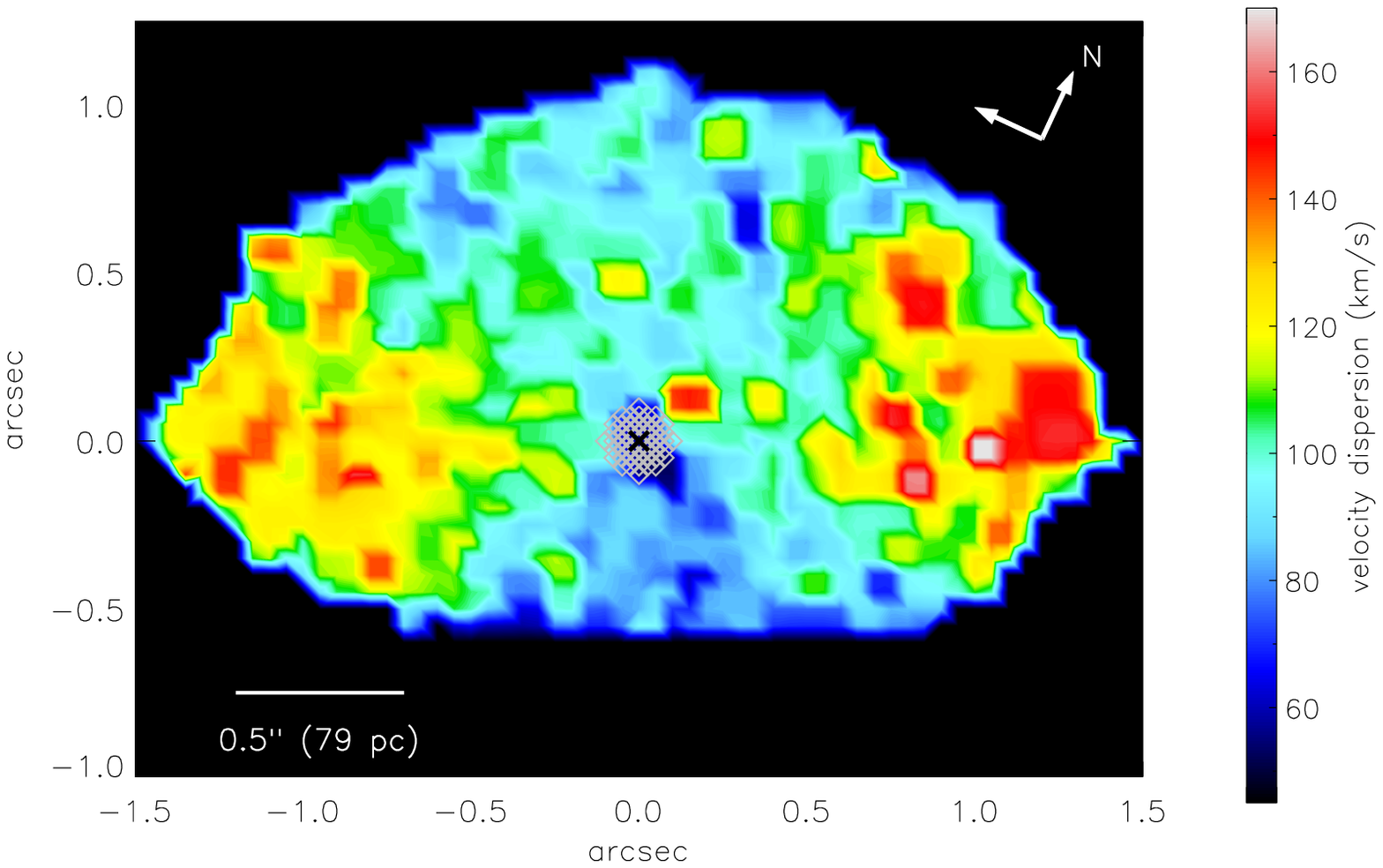,width=0.6\linewidth,clip=} 
\caption[ppxf_fitting]{Two-Dimensional map of line of sight velocity and velocity dispersion from the spectral fitting with pPXF. The orientation is the same as in the right panel of Fig.~1. The white bins were masked out due to the high non-stellar contribution and the black cross indicates the AGN position. Top: Line of sight velocity map. Positive and negative velocities correspond to movement away and towards the observer respectively. The values were corrected for the systemic velocity of the galaxy but not for the inclination ($\sim 55$ degrees). To note that the values are more uncertain in the central pixels due to the presence of a high AGN continuum. In the central regions the stars rotate in the clockwise direction while in the outer regions they rotate in the opposite direction (seen as a change of velocity direction in the top left and top right regions of the map). Bottom: Map of velocity dispersion. The velocity dispersion is higher in the outer regions, decreasing when approaching the nuclear region. The higher values of dispersion occur in a plane, showing little vertical extension. The values of velocity dispersion have been corrected for the instrumental broadening.}
\label{vel_field}
\end{figure*}

We first run pPXF in the integrated field-of-view, excluding the central regions dominated by the AGN continuum to increase the signal-to-noise ratio. We plot the results, which correspond to our maximum signal-to-noise ratio for these data (S/N $\sim 10$ for the deepest absorption features or S/N $\sim 150$ in relation to the total continuum) (Fig.~\ref{fit}). Our stellar templates provide a good fit to the data, with the main contributions of the stellar templates of a K5 III and a M0 III star. 
We removed an absorption feature that the stellar templates were not able to fit, it coincides with the wavelength of the Fe I stellar absorption and a telluric feature. The fact that we are not able to fit it could be due to telluric residuals and/or a metallicity higher than the metallicity of our templates (which do not include super-metallicity stars).

In Fig.~\ref{region_spectra} we show the spectral data and spectral fitting at four different regions of the field-of-view (numbered boxes in the top panel). The spectra plotted in the bottom panels, show the integrated spectrum (black), best-fit model (red) and the residuals (grey), for each of the four different regions. It is clear that the stellar features are deeper close to the major axis of the galaxy (regions 1 and 3), where we also see the stronger stellar continuum.

We follow the same procedure as described above but now across the entire field-of-view on a spaxel-by-spaxel basis, fitting the wavelength range 
$\lambda_{\rm rest}$ (1.57 - 1.716) $\micron$, to obtain a map of stellar velocity and velocity dispersion. We show the results in Fig.~\ref{vel_field}. The systemic velocity is determined using the method described in Appendix C of \cite{krajnovic06}, using the velocity maps of the central $r > 1''$. We obtain best fit values of $V_{\rm sys} = 2396$ km/s and a kinematic PA of 112$\pm$16 degrees. 
The velocity is corrected for the galaxy's systemic velocity, but not for the inclination. From work of \cite{lauberts82} compiled in \cite{dezotti85}, the axis ratio for this galaxy is $b/a = 0.58$. The axis ratio of our stellar continuum is consistent with the value quoted above, indicating that the axis ratio at small and large scales is the same. From the axis ratio of the stellar continuum (and assuming a flat geometry), we derive an inclination of $\sim 55$ degrees. Our rotation velocities could be up to $49$ km/s, instead of the observed $\sim40$ km/s. As described in Section~\ref{sec:extracting} the velocity dispersion obtained already takes into account the instrumental broadening. In the central regions (r $<$ 0${''}$.1) around the black hole, the non-stellar continuum is very strong, and the relative intensity of the stellar absorption lines is low, which increases the error in determining the velocity dispersion. For this reason we masked out the inner r $<$ 0${''}$.1 (white bins in Fig. \ref{vel_field}). The errors in the parameters are measured using a Monte Carlo approach. We generate a random set of 100 spectra with the same noise properties (S/N $\sim 5$) as our galaxy spectra and fit them with pPXF. The starting values and the wavelength limits of the fit are changed randomly as well. The final 1$\sigma$ deviations in each parameter are taken as our errors for a typical spectrum with S/N $\sim$ 5 (which corresponds to the S/N in each bin). We find errors of $\pm4.5$ km/s for the mean line-of-sight velocity and $\pm5.1$ km/s for the velocity dispersion.
\subsubsection*{The counter-rotating core}
The results show a low rotational velocity when compared with the velocity dispersion. The zero point in the velocity map corresponds to the systemic velocity of the galaxy, negative velocities are associated with blue-shifted absorption lines and positive velocities with redshifted absorption lines. The central $\sim 1.4$ arcsec radius region shows the evidence of two distinct kinematic components. The inner and outer regions of our map indicate two different and counter-rotating components, with the $r < 0{''}.8 \sim 125$ pc region rotating clockwise and the $r > 0{''}.8$ rotating counter-clockwise. The two kinematic components have similar absolute velocity values ($40 - 50$ km/s). The rotation of the counter rotating core is seen clearly in the figure. The outer region is only beginning at the limits of our analysed region of the field-of-view - corresponding to the region above the flux threshold. It is visible in the top right (yellow/red zone) and top left (blue zone) of the velocity map. The velocity dispersion map presents values that decrease inwards and are flat at the centre. It has values of $\sim130$ km/s in the outer regions $r > 0{''}.8$ but drops by $\sim30$ km/s in the inner regions, which coincides with the counter-rotating core. The dispersion values are in general higher than the line-of-sight velocity, which suggests that the galaxy core is not rotationally supported.

Although we do not use the $h3$ distribution in our analysis due to the low signal-to-noise ratio, we did a test to check if it was consistent with the velocity map of the top panel of Fig.~\ref{vel_field}. A value of $h3$ different from zero indicates that the line-of-sight velocity distribution profile deviates from a Gaussian distribution. The value of $h3$ is related with the skewness, and measures the asymmetry in the distribution. As expected, the $h3$ map is very clearly anti-correlated with the line-of-sight velocity distribution for $r < 0{''}.8$ and in the upper regions of the map where the velocity changes sign (although not as clearly due to the higher noise level in those pixels).  

The general kinematic properties mentioned above, indicate that we are most likely in the presence of a kinematically distinct core. These type of systems are fairly common in ellipticals but counter-rotating systems are rarer ($< 10$ per cent) in S0 type galaxies \citep{kuijken96,krajnovic11,bois11}.
The drop in the velocity dispersion occurs for $r < 0{''}.8$ and is possibly related with the counter-rotating core. This phenomenon has been observed for another S0 galaxy with a kinematically-distinct core, NGC 7332, where the velocity dispersion increases from $r\sim10''$ or 1 kpc towards the centre of the galaxy, but drops by $\sim 10$ km/s when in the region of the counter-rotating core \citep{falcon-barroso04}. Our data cover a smaller radii ($r_{\rm max}\sim1{''}.4$ or 200 pc) than the work of \cite{falcon-barroso04}, which allow us to measure the drop in velocity dispersion in the inner regions, but does not permit us to evaluate if the velocity dispersion decreases again as we move to larger radii.
The dynamically decoupled core could be the result of an inflow of gas into the central regions of the galaxy via, for example, a past minor merger event. For our target, the majority of the stellar population is old and fairly homogeneous, not showing any variations with radius \citep{boisson02}. There is nevertheless indication of a series of previous star formation bursts \citep{bonatto00}. The evolution for this galaxy is not expected to be through major mergers, since only about $2 - 2.5$ per cent of S0 galaxies are expected to have had a major merger in their past \citep{parry09}. On the other hand, secular evolution is expected to dominate the evolution of Narrow Line Seyfert 1 galaxies such as MCG--6-30-15 (e.g. \citealt{orbandexivry11}).

\subsection{Gas kinematics}
\begin{figure}
\begin{centering}
\includegraphics[width=1.0\columnwidth]{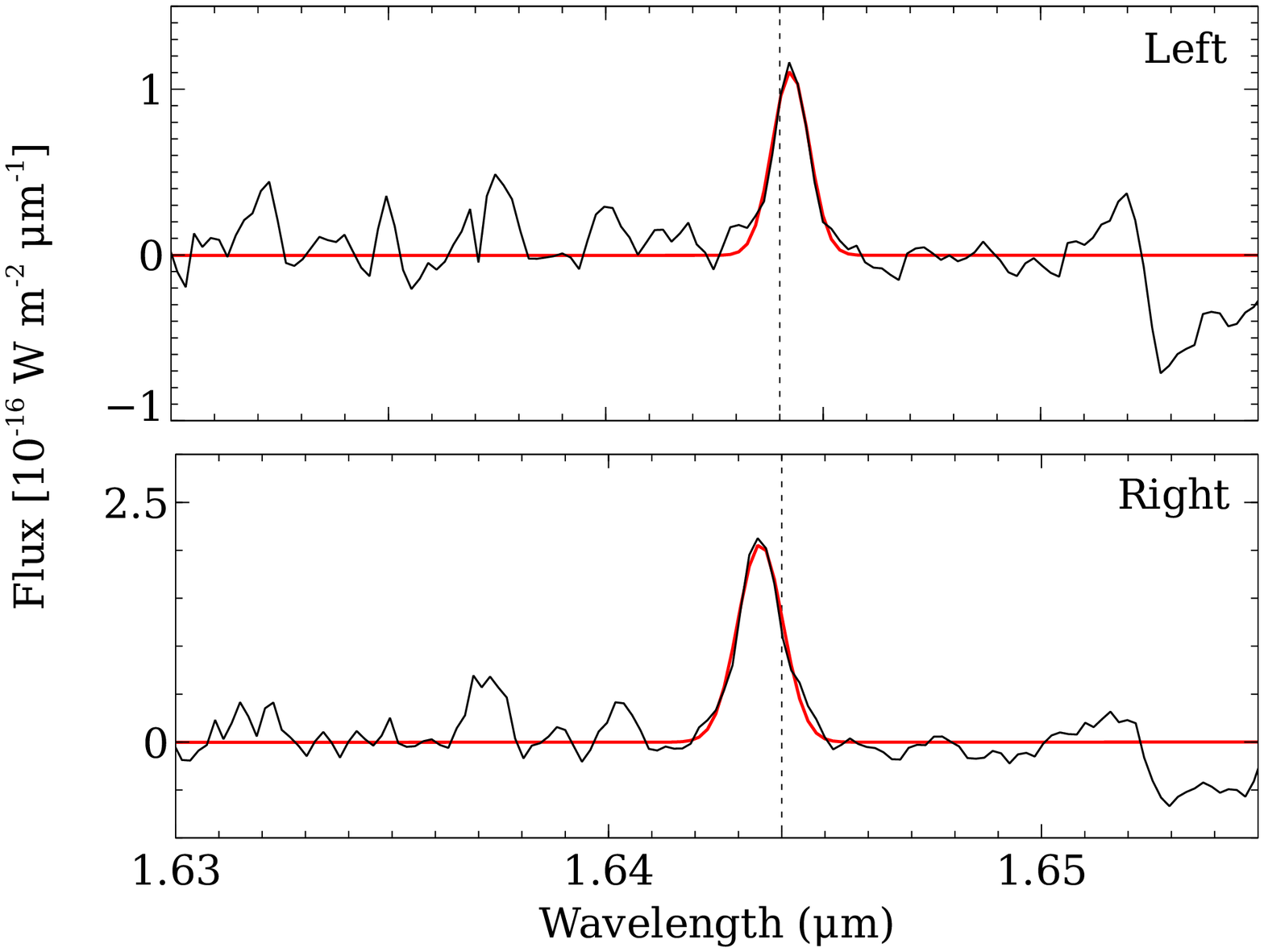}
\caption [Fe II]{[Fe II] emission integrated in two regions (10$\times$10 pixel each) of the field-of-view and plotted as a function of the rest-frame wavelength. East (top panel) and West (bottom panel) of the nucleus, along the galaxy major axis. The gas shows a different velocity in both regions, as can be seen from the wavelength shift in the plot. The integrated Western region shows a larger velocity dispersion and intensity than the emission from the Eastern region. The specific values are given in the text.}
\label{iron_line}
\end{centering}
\end{figure}
\begin{figure}
\centering
\begin{tabular}{ccc}
\epsfig{file=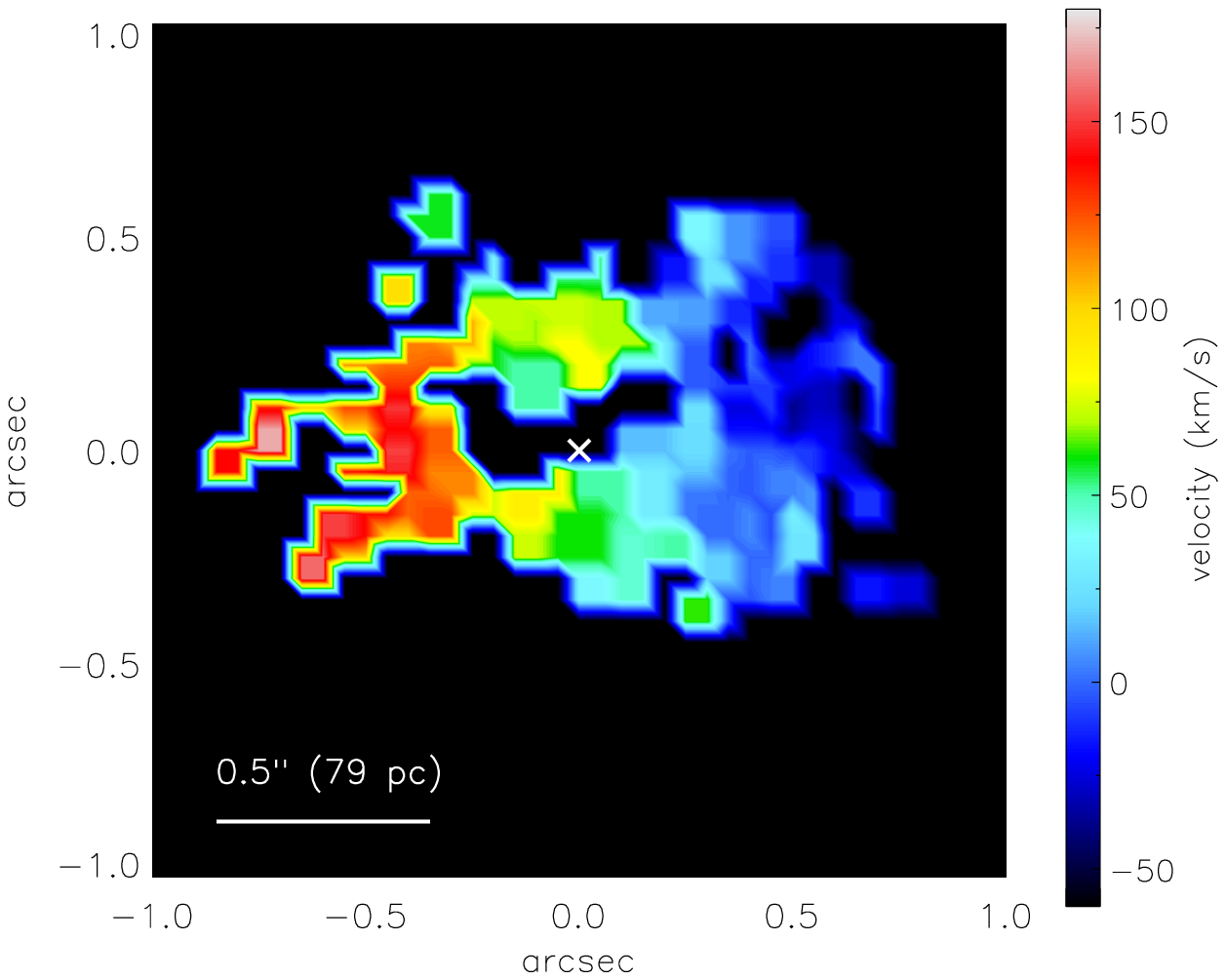,width=1.1\columnwidth,clip=} \\[0.1cm]
\epsfig{file=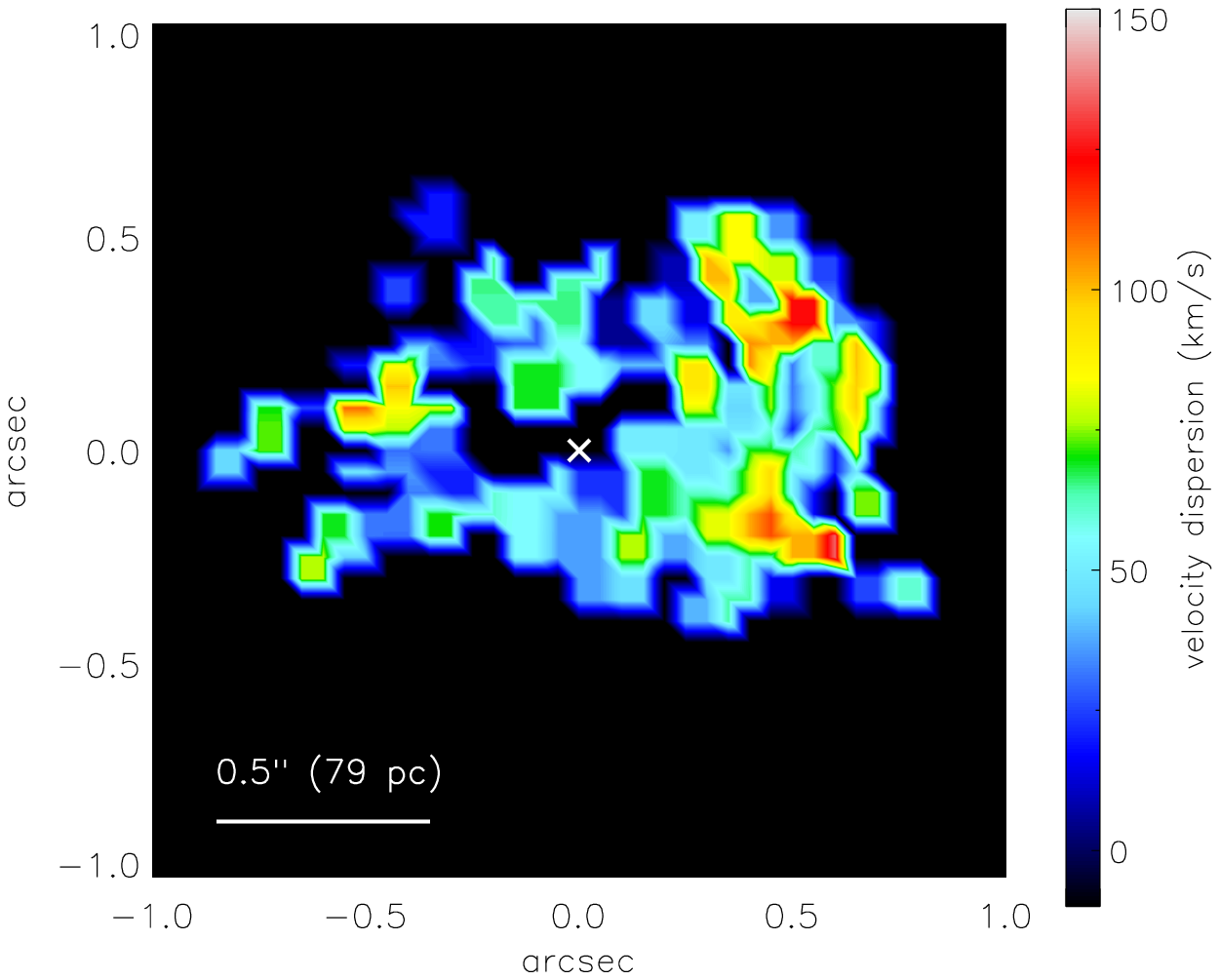,width=1.1\columnwidth,clip=} \\[0.1cm]
\epsfig{file=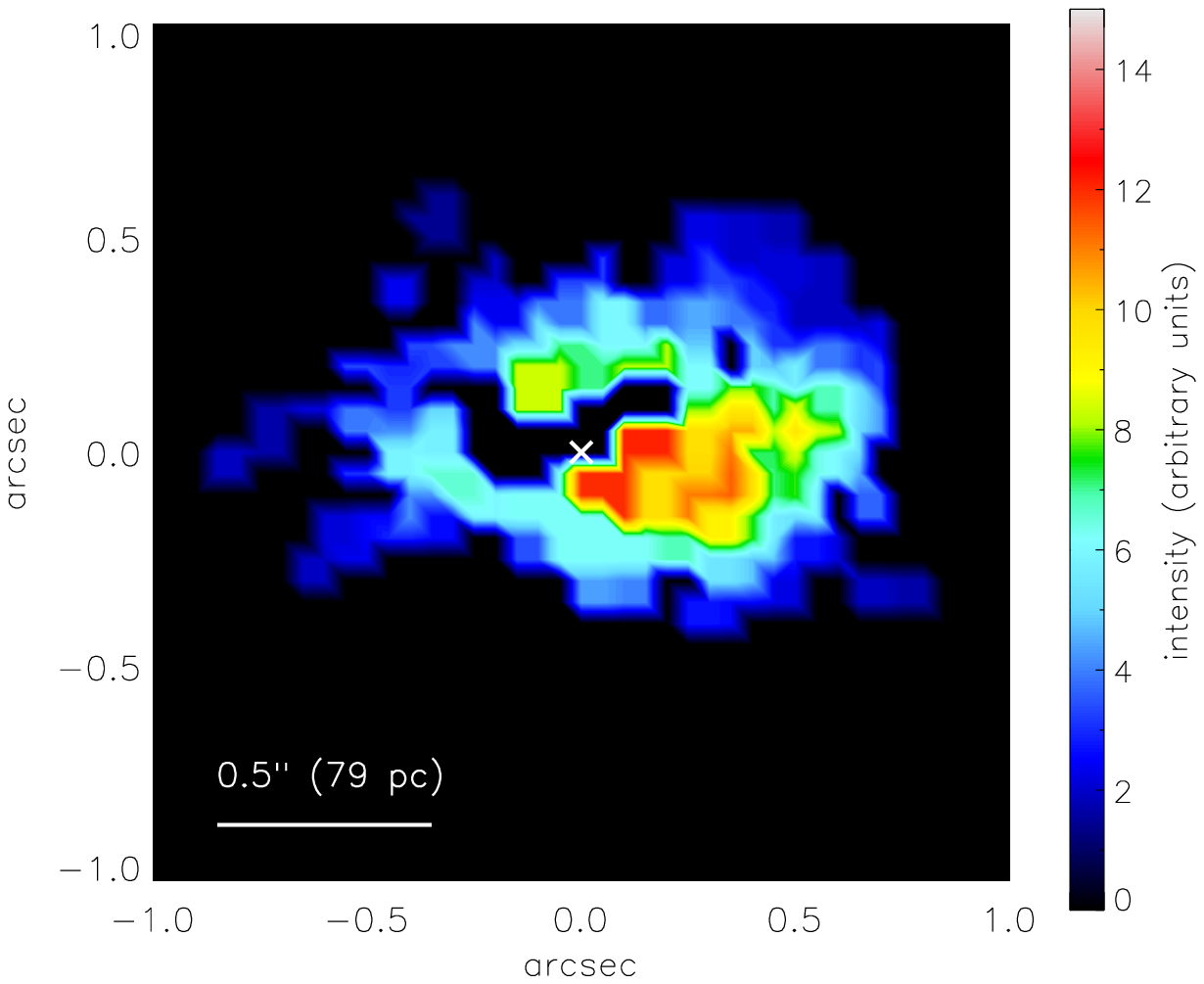,width=1.1\columnwidth,clip=}
\end{tabular}
\caption[Iron]{Panels showing the [Fe II] $\lambda = 1.644 \micron$ emission properties of MCG--6-30-15 obtained by fitting the line with MPFIT. The white cross marks the position of the AGN and the map orientation is the same as for the right panel of Fig. \ref{hst_sinfoni}. The regions referred to in the text as to the East and to the West of the nucleus are measured along the major axis. Bins with S/N $< 3$ where excluded and bins for which the velocity dispersion is lower or similar to the instrumental broadening were also excluded (the region surrounding the black hole position did not obey this condition and hence was masked out). Top panel: Velocity corrected for the systemic velocity of the galaxy in km/s. Middle panel: Velocity dispersion corrected for the instrumental broadening in km/s. Bottom panel: Line intensity. The line intensity is higher on the West side of the nucleus than on the East side. The velocity shows an asymmetry, with higher absolute values on the East side of the nucleus than on the West. The two hypothesis for this [Fe II] emission (AGN outflow and most likely supernova induced shocks) are discussed in the text.}
\label{iron}
\end{figure}
When subtracting the stellar and AGN continuum from our data cube using the results from pPXF, we are left with the gas emission spectra at each spatial location. The $\lambda = 1.644\thinspace\micron$ forbidden emission line of [Fe II] is the strongest emission line, and clearly observed by eye in the spectra.  
In Fig.~\ref{iron_line} we plot the fit to the emission line integrated in two regions of the same size (10 x 10 pixels) offset by 0${''}$.5 from the AGN position on the East and on the West side of the nucleus. As we can see from the plot, the intensity is larger on the West than on the East side. The line velocity is also different, the West side is blueshifted ($-8$ km/s) and the East side is redshifted ($+129$ km/s). The lines show a broadening of $49$ km/s and $74$ km/s on the East and West side of the nucleus respectively.
In the individual pixels, the S/N of the line is not as strong. To get the spatial distribution of [Fe II], we fit a Gaussian to this line and fit the continuum using regions on the left and right of the line using the algorithm MPFIT (\citealt{markwardt09}, \citealt{more78}). We require a signal-to-noise ratio S/N $>3$ in relation to the RMS scatter of the spectrum, a condition that in general holds in the central regions of the galaxy and coincides with the kinematically-distinct core.
We show the two-dimensional line properties in the three panels of Fig. \ref{iron}, the top panel shows the velocity offset, the middle panel the velocity dispersion and the bottom panel the line intensity. In these plots, the velocities are measured in relation to the systemic velocity of the galaxy. The instrumental broadening is subtracted in quadrature using the dispersion of an unblended sky emission line close to the [Fe II] line emission. We excluded regions that showed velocity dispersion values similar to the instrumental broadening.

The lines show a gradient from the minimum values blueshifted by $\sim-30$km/s on the West side of the nucleus to the larger redshifted values ($+120$ km/s) on the East side. We observe an asymmetry in the intensity map, with larger values on the West side of the nucleus than on the East.
The velocity dispersion is harder to interpret, it is higher on the West side of the nucleus ($\sim60$ km/s) but decreases around the AGN position. It seems to increase again on the East side although the signal-to-noise ratio there is not as good.

\subsubsection*{The origin of the [Fe II] emission}
Since the signal in the data is weak, we cannot say with certainty what causes this [Fe II] emission, but we can discuss the two possibilities: an AGN outflow or supernova shocks.
The intensity of [Fe II] is asymmetric, with stronger emission on the West side of the nucleus. The velocity map for [Fe II] shows that the gas has a distribution along the main axis of the galaxy and the values show a deviation from the rotation velocity of the stars. The distribution of [O III] $5007$ \AA\thinspace\thinspace emission using HST observations (\citealt{schmitt03} and \citealt{bennert06}) show an elongation along the major axis of the galaxy, as we see in the [Fe II] distribution. The [O III] emission also shows an asymmetric distribution but on a larger scale than the field-of-view of our observations.
The fact that we see a velocity structure distinct from the stellar rotation pattern, and an elongation similar to the [O III] emission, could indicate that we are in the presence of an AGN driven outflow, with [Fe II] and [O III] emission tracing the same geometry (e.g. \citealt{muller-sanchez11}). The gas kinematics could be consistent with emission from a cone that is outflowing away from us.

There are nevertheless some caveats in this hypothesis. The velocity offset from the rotational velocity is not very large, $\sim10$ km/s on the West side and $\sim 80$ km/s on the East side of the nucleus (where the S/N is lower), and could be due to uncertainties in the velocity measurements. The transition in velocity as it moves across the nucleus is fairly smooth and resembles a rotation pattern.
Our data samples a much smaller spatial scale than the [O III] maps, and we cannot check if the asymmetry of the [Fe II] intensity is observed at the larger scales as well. We are limited to the small scale of our observations and the low signal-to-noise. With new data on a larger scale, it would be possible to investigate if the asymmetry in the gas emission holds for larger radii. It is also known that at the very small scales probed by the X-ray observations of this source, there are spectroscopic signs of a warm absorber (e.g. \citealt{reynolds97}, \citealt{ballantyne03}), and hence a form of outflow. It would be interesting to investigate this topic further to determine if there is any connection between the accretion disc scale outflow and the tens of parsecs of the [Fe II] emission.

The [Fe II] emission is excited by electron collisions and is a good tracer of shocks. In the case of MCG--6-30-15, the extended emission could be caused by nuclear mass outflow shocks with ambient clouds, or by supernova-driven shocks \citep{mouri00}.
The spatial distribution of the [Fe II] coincides with the region where we see the strongest signal-to-noise in the stellar absorption features. This and the arguments presented above, imply that most likely the [Fe II] emission is due to shock fronts from supernova remnants.
These shocks destroy the dust grains which allows for a higher abundance of iron in the gas phase. The iron is then ionised by the interstellar medium.
The [Fe II] line is therefore a good indicator for the presence of supernova remnants and a tracer of the supernova rate (\citealt{moorwood&oliva88, colina93, rosenberg12}). 

In this case, the [Fe II] emission can be used to constrain the supernova rate and investigate the properties of the stellar population of MCG--6-30-15, as we will discuss in Section~\ref{sec:stellarpop}.

\subsection{Central stellar population}
\label{sec:stellarpop}
From the observations and results of the previous sections, we can investigate the properties of the stellar population of MCG--6-30-15.
Assuming that the supernova shocks were the main excitation mechanism for [Fe II], we can determine the supernova rate (SNR) based on the [Fe II] 1.644 $\micron$ line flux. We measure a flux (integrated in the $r<0{''}.8$ region) of F$_{[Fe II] 1.644} = 8.2 \times 10^{-16}$erg s$^{-1}$ cm$^{-2}$ which corresponds to a luminosity of $L_{[Fe II] 1.644} = 1.2 \times 10^{38}$erg s$^{-1}$. This [Fe II] is not particularly strong, its flux is comparable to the lower values found for narrow line Seyfert 1 galaxies in the sample of \cite{rodriguez-ardila04}.
Using the equations from \cite{rosenberg12}:
\begin{eqnarray}
\log \frac{\rm {SNR}}{\rm [yr^{-1}]} = (0.89 \pm 0.2)\times \log \frac{L_{[Fe II] 1.26}}{[\rm erg s^{-1}]} - (36.19 \pm 0.9)
\end{eqnarray}
and the theoretical ratio [Fe II] 1.64 $\micron$/[Fe II] 1.26 $\micron$ = $0.7646$ from \cite{nussbaumer&storey88}, we calculate a supernova rate of SNR = $6.6\times10^{-3}$ yr$^{-1}$.

We used the evolutionary synthesis code STARS \citep{sternberg98, thornley00, sternberg03} to follow the evolution of a stellar population as in \citealt{forster-schreiber03,davies05} and \citealt{daviesetal07}. In the code, the star formation declines exponentially with a characteristic timescale, which we assume to be 10 Myr, adopted from \cite{daviesetal07}. Assuming a Salpeter Initial Mass Function and this star formation timescale, STARS calculates the distribution of stars in function of age. The code results are scaled to the case in study by comparing the \emph{K}-band luminosity in the model with our observed \emph{H}-band luminosity. The conversion between \emph{H} and \emph{K}-band luminosities is done using the magnitude relation \emph{H} - \emph{K} = 0.15 which is very weakly dependent on the stellar age. The code calculates parameters of interest for our investigation, such as the supernova rate, \emph{K}-band luminosity, mass loss rate and total stellar mass at any time.
Using the relation between predicted SNR and the stellar age from the code, we determine the age of the starburst that produced the SNR observed. From our SNR we obtain an age of $~6.5\times10^{7}$ yr. We note that although two values for the age are consistent with the supernova rate measured, the alternative younger age ($3\times10^{6}$ yr) is inconsistent with the presence of deep absorption features that indicate the presence of late-type stars.
\begin{figure}
\begin{centering}
\includegraphics[width=0.8\columnwidth]{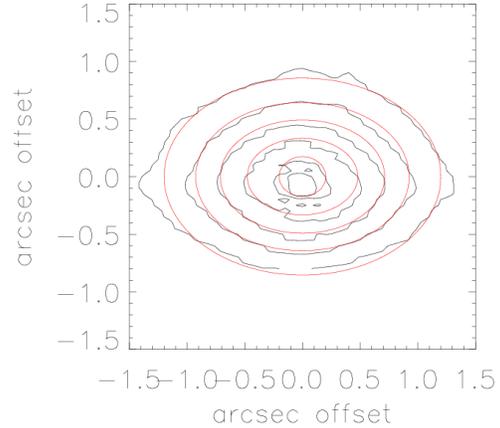}
\caption [mge]{Multi-Gaussian Expansion (MGE) modelling of the galaxy surface brightness. Data (black contours) compared with the MGE model (red contours).}
\label{mge}
\end{centering}
\end{figure}
With a stellar age of $~6.5\times10^{7}$ yr we obtain from the model a mass-to-light ratio of $M/L_{\rm H} = 0.98$ M$_{\odot}$/L$_{\odot}$.
From the model we also know the mass loss rate, which can be compared with the mass accreted by the black hole. The stronger stellar and [Fe II] emission coincide with the counter-rotating core. This type of structure is thought to be formed by an inflow of new gas via, for example, a minor merger that generates a stellar population with different properties in the nucleus of the galaxy. New gas may have generated the stellar population we observe and model with STARS, but can also have fuelled the AGN. From the bolometric luminosity ($\sim 1.3\times10^{44}$ erg s$^{-1}$; \citealt{vasudevan09}), and assuming an efficiency of $\epsilon = 0.1$ we can obtain an estimate for the mass accretion rate $\dot{M} = L_{\rm bol}/\epsilon c^2$. For this AGN the mass accretion rate is around $0.022$ M$_{\odot}$yr$^{-1}$, which gives, assuming a constant mass accretion rate during the age of the starburst, a total mass accreted of:

\begin{eqnarray}
M_{\rm tot} = 6.5\times10^{7}\times 0.022 = 1.4\times10^{6} M_{\odot}
\end{eqnarray}
The model-derived total mass used to form stars is $2\times10^{8}$ M$_{\odot}$. This means that of the new inflow of gas, only around $1$ per cent of the mass used to form stars has been used to fuel the supermassive black hole.

Another alternative would be for the mass loss by the stars to be the main source of fuel for the AGN. The mass loss rate for this population is $1$ M$_{\odot}$yr$^{-1}$, which is larger than the mass needed to maintain the AGN activity, providing that the process of transferring ejected mass from stars to the black hole is reasonably efficient. 
In either case, it is, nevertheless, still uncertain how this mass can move inwards to $< 1$ pc scales to fuel the black hole.

As a cross check, we investigated the influence of considering a larger characteristic timescale in our STARS models. We repeated the above calculations for a timescale of 100 Myr. The conversion between the supernova rate measured and the age from the STARS model will be different, and will provide an upper limit for the age ($2.5\times10^{8}$ yr). The mass-to-light ratio would be M/L$_{\rm H}$ = $1.3$M$_{\odot}/$L$_{\odot}$, and a mass loss rate of $\dot{M}_{\rm loss} = 2.1$M$_{\odot}$/yr. The conclusions remain unchanged, the mass loss rate of the stars would in principle be enough to fuel the black hole, or, if the gas that created the stars also fuelled the black hole, the fraction of gas for the AGN compared with gas used to form stars is again around $1$ per cent.

In summary, it is possible that the counter-rotating core is associated with new inflow of gas which fuelled the formation of a new stellar population with distinct stellar kinematics. The [Fe II] emission traces the supernova and is observed in the inner regions where the counter-rotating core is. We used the measured [Fe II] flux and the STARS code to learn more about the episode of star formation and its relation with the black hole fuelling.
The starburst has an age of approximately $6.5\times10^{7}$ yr, formed $2\times10^{8}$M$_{\odot}$ stars and presents a mass loss rate of $1$ M$_{\odot}$yr$^{-1}$. If the AGN activity is fuelled by ejections from the stars, we would need a mass loss comparable to the mass accretion rate of the black hole, which is $0.022$ M$_{\odot}$yr$^{-1}$. The mass loss rate obtained using STARS would be enough to fuel the black hole activity, providing that the process is efficient.
On the other hand, if the new inflow of gas was responsible for the star formation and for providing the fuel to the black hole, we would need an amount of gas capable of powering the AGN for $6.5\times10^{7}$ yr. Considering that the black hole was always active and at a constant mass accretion rate, the total initial gas mass needed for black hole fuelling would be $1.4\times10^{6}$M$_{\odot}$ which is roughly $1$ per cent of the total mass used to form stars.
These calculations provide approximate values only, but are useful aids to understanding the general mass budget in the central region of the galaxy. 
\begin{figure*}
\begin{centering}
\hspace{-0.25cm}\includegraphics[width=2.1\columnwidth]{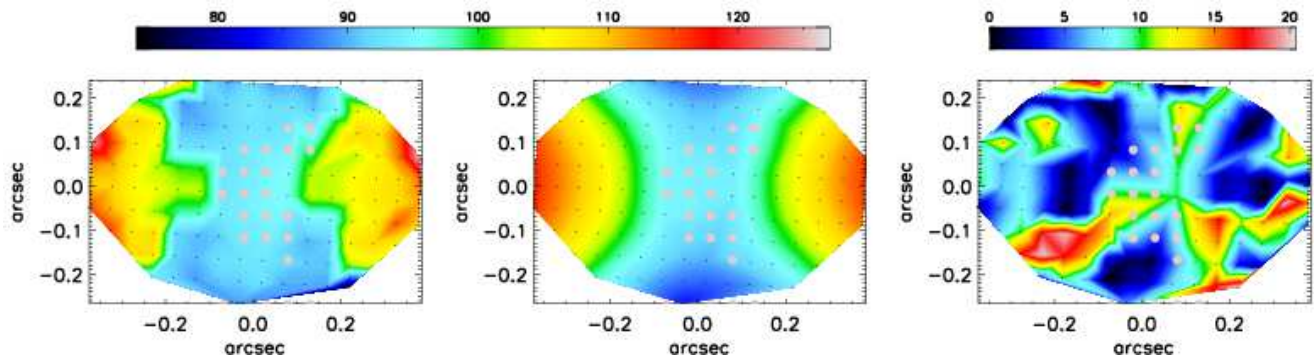}
\caption {Comparison between the symmetrised observed (left) and the JAM predicted $V_{\rm rms}$ (center) for the best-fit set of parameters. The absolute values of the residuals are shown in the right panel: $|V_{\rm rms}^{\rm JAM}-V_{\rm rms}|$. The colour-bar is in km/s.}
\label{jam_bestfit}
\end{centering}
\end{figure*}

\subsection{Black hole mass}
The spatial zone where the black hole potential dominates over the galaxy potential, the black hole sphere of influence, is given by the radius from the black hole: $R = GM_{\rm BH}/\sigma^{2}$. For MCG--6-30-15, using the upper limit values from \cite{mchardy05}, the radius of influence is $R\sim3$ pc or $0''.02$. This scale is not resolved in our data, since our spaxel size is $0''.05$ and the PSF has a FWHM $\sim 0''.1$. The galaxy potential will contribute significantly to the measured dynamics, which makes it harder to determine the black hole mass directly from the central stellar kinematics, without doing a multi-component dynamical model.
Previously, the black hole mass for this galaxy was determined from the black hole relation with the host galaxy (M$_{\rm BH}$ - $\sigma$). The effective radius in this galaxy is $R_{e}\sim 9''$ \citep{boisson02}. \cite{mchardy05} use the relation from \cite{ferrarese02} defined at a radius of $R_{e}/8 \sim 1{''}.12$ and velocity dispersion from long slit spectroscopy to obtain a black hole mass of (3 - 6) $\times 10^{6}$ M$_{\odot}$.
We determine the velocity dispersion from a pseudo slit with the same size as the one used by \cite{mchardy05}. The integrated value is $\sigma=109$ km/s, which is slightly higher than that obtained by \cite{mchardy05} of $93.5\pm 8.5$ km s$^{-1}$. The values agree marginally within the errors. The difference could also be related with the distinct stellar population that we probe using infrared observations compared with the one observed in the optical. The method of inferring the black hole mass based on the velocity dispersion is known to be subject to large uncertainties. In the case of MCG--6-30-15, the decoupled kinematic components in the centre of the galaxy could also affect the velocity dispersion measurement.

As a first approach, we use the velocity and velocity dispersion at the closest resolved region around the black hole to determine an upper limit for the black hole mass. The enclosed mass within radius $R$ is, from virial arguments, $M_{\rm enc} = (v^{2} + 3\times \sigma^{2}) R/G$. We determine the velocity dispersion from the integrated spectra in the inner $R < 0{''}.2$ region around the black hole, excluding the $R < 0{''}.1$ due to the high non-stellar continuum. The value obtained is $\sigma = 89 \pm 8$ km/s, and the typical velocity is $10$ km/s which gives an enclosed mass of $M_{\rm enc} = 1.7 \times 10^{8}$M$_{\odot}$. From the calculations in Section \ref{sec:stellarpop}, we can obtain a lower limit estimate (because it is only the mass of the young stellar population) for the stellar mass using the observed luminosity in the \emph{H}-band within a radius of $R < 0{''}.2$ and the mass-to-light ratio determined from STARS: M$_{\rm stel} = 2.4 \times 10^{7}$M$_{\odot}$. An upper limit for the black hole mass can be found by subtracting the stellar mass from the enclosed dynamical mass: M$_{\rm BH} < 1.5 \times 10^{8}$M$_{\odot}$.

Alternatively, we try to constrain the mass of the black hole and the dynamics of the galaxy with the Jeans Anisotropic Model (JAM) method of \citealt{cappellari08}. This model generalises the asymmetric Jeans equations to the case of anisotropy by including an anisotropy parameter $\beta_{z} = 1 - \bar{\sigma_z^{2}}/\bar{\sigma_r^{2}}$. 
The model we use assumes axisymmetric geometry and takes as input the galaxy surface brightness to generate a prediction for the second velocity moments ($V_{\rm rms} = \sqrt{V^{2}+\sigma^{2}}$), for a combination of physical parameters: inclination, black hole mass, mass-to-light ratio and $\beta$. The output model is then compared, and adjusted if necessary, to the observed kinematic map of the galaxy.
The galaxy surface brightness is given as a combination of gaussians, parametrised by a Multi-Gaussian Expansion (MGE) fitting software developed by \citealt{cappellari02}.
We start by doing a multi-Gaussian expansion of our \emph{H}-band map, constrained to a region of $r < 0''.8$ to exclude the outer counter-rotating component.
The surface brightness can be described by three gaussians, we plot the resulting contours in Fig.~\ref{mge}.
The gaussian parameters are converted to the units required by JAM using the guidelines from the JAM code release by Michelle Cappellari. 
\begin{figure*}
\begin{centering}
\includegraphics[width=2.0\columnwidth]{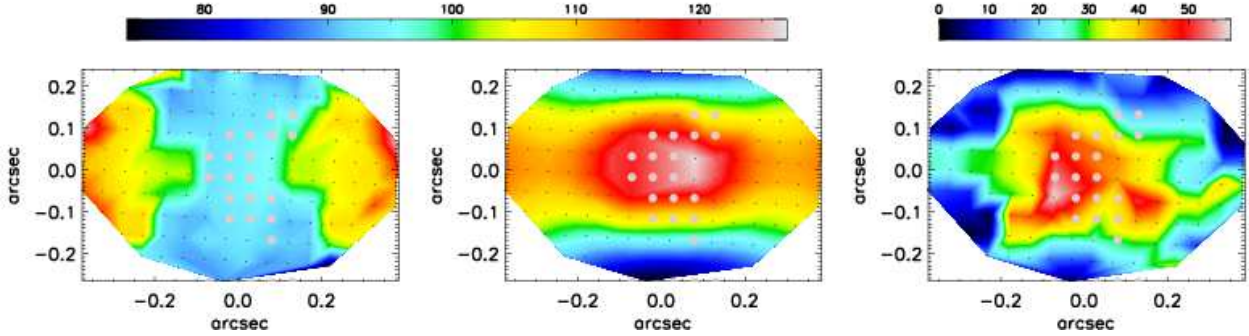}
\caption {Comparison between the symmetrised observed (left) and the JAM predicted $V_{\rm rms}$ (centre) corresponding to a black hole mass upper limit of $1.5 \times 10^8$M$_{\odot}$ and the remaining parameters similar to Fig.~\ref{jam_bestfit}. The absolute value of the residuals are shown in the right panel: $|V_{\rm rms}^{\rm JAM}-V_{\rm rms}|$. The colour-bar is in km/s.}
\label{jam_high}
\end{centering}
\end{figure*}
We are conservative in the number of bins we fit, and exclude the central pixels that show the largest errors in the velocity dispersion determination. The errors in velocity and velocity dispersion at each bin were determined using the same method as in Section~\ref{sec:stellarkin}. We run 100 Monte Carlo simulations for each bin, setting the error statistics to be the same as the data at that position. The result is a two-dimensional map of velocity and velocity dispersion errors that are taken as an optional input map for the Jeans modelling.

The model output and parameters are affected by the area in which the modelling is done. A change in the black hole mass affects mainly the central bins and does not change much the $V_{\rm rms}$ map in the outer regions. The lower limit for the black hole mass is difficult to determine, since the $V_{\rm rms}$ is consistent with a null black hole mass (the $\chi^{2}$ of the model is not very sensitive to variations in the mass below $\sim 10^{7}$M$_{\odot}$). The black hole mass value and the mass limit will depend on the area where we calculate the $\chi^{2}$. For a larger area, and hence more degrees of freedom, the reduced $\chi^{2}$ will increase more slowly. We use the region $r < 0''.4$ to limit the analysis to the area of higher black hole influence and at the same time to have enough bins for the calculation. The least-squares fitting routine MPFIT is used to search the parameter space for $M_{\rm BH}$, inclination, $\beta$ and M/L and determine the best fit parameters. We then fix all parameters, except the black hole mass, to their best fit values and determine an upper limit for the black hole mass based on the $\chi^{2}-\chi_{\rm min}^{2}$ distribution. 
The absolute values of the parameters in the fit are hard to constrain, due to possible degeneracies and the effect of the radius of the area analysed. The degeneracy between the inclination and $\beta_{z}$ parameters for example, could be removed with observationally motivated constraints (e.g. \citealt{cappellari08}). Unfortunately we do not have external information on the inclination of the galaxy. Following the arguments presented above, we focus on using the model to determine an upper limit for the black hole mass.

The 1$\sigma$ confidence limit gives an upper limit for the black hole mass of $M_{\rm BH} < 6\times 10^{7}$M$_{\odot}$. The best-fit black hole mass is $4 \times 10^{6}$M$_{\odot}$ and a comparison between the data and model is shown in Fig.~\ref{jam_bestfit}. However, this best-fit value should be taken with caution due to the arguments discussed above; the upper limit provides a stronger constraint on the black hole mass.

An increase in the black hole mass causes the $V_{\rm rms}$ of the previous analysis to increase in the inner regions of the galaxy. In Fig.~\ref{jam_high} we plot an example of this effect, by fixing the parameters to be the same as in Fig.~\ref{jam_bestfit} but increasing the black hole mass to the upper limit determined from the simple calculation with the integrated velocity dispersion: $1.5\times 10^{8}$M$_{\odot}$. It is clear that the increase in $V_{\rm rms}$ expected from models with high black hole mass is not observed in the data, which confirms qualitatively the value determined as an upper limit.
This upper limit was not derived based on the M-$\sigma$ correlation, and hence provides an independent measurement. Nevertheless, if we take the upper limit for the black hole mass ($6\times10^{7}$M$_{\odot}$) and the integrated velocity dispersion ($\sigma=109$ km/s) we measure, the location of MCG--6-30-15 in the M$_{\rm BH} - \sigma$ plot is in agreement with the M$_{\rm BH} - \sigma$ relation (e.g. \citealt{gultekin09}).
We conclude that the mass of the black hole in MCG--6-30-15 is lower than $6\times10^{7}$M$_{\odot}$ which is in agreement with the previous estimates.

For a bolometric luminosity of $~1.3\times10^{44}$ erg s$^{-1}$ \citep{vasudevan09}, we obtain a lower limit for the Eddington ratio of $\lambda > 0.02$.

\section{Conclusions}
\label{sec:conclusions}
In this work we studied for the first time the inner $\sim 470$ pc of the galaxy MCG--6-30-15 using integral field spectroscopy. We were able to remove the AGN broad hydrogen Brackett emission lines which were dominating the spectra in the central region of the field-of-view, and measure the properties of the stellar absorption lines. 
The stellar kinematics of this galaxy can be characterised by a low rotational velocity ($\sim 40$ km/s) compared with the velocity dispersion ($80 - 140$ km/s). The velocity dispersion is higher at larger radii and close to the major axis and decreases when approaching the position of the black hole. There is a change in the direction of the stellar rotation when comparing the central core ($r < 0{''}.8$) with the outer regions. We argue that we are in the presence of a galaxy with a counter-rotating core, due to the rotation curve and the observed drop in velocity dispersion at small radii.
The gas dynamics traced by the [Fe II] emission line show an asymmetric distribution in the inner $r < 0{''}.8$ arcsec, with a higher intensity and larger velocity dispersion on the West side of the nucleus. It also shows a velocity gradient with blueshifted velocities of $-30$ km/s on the West side and $+120$ km/s on the East side of the nucleus. 

The quality of the data does not allow us to exclude the possibility of an outflow, but due to the smooth velocity curve and its spatial distribution we argue that the [Fe II] has been excited by supernova shocks. In this scenario we use a model to reproduce the supernova rate inferred and determine the star formation history. If the counter-rotating core is a result of a recent inflow of gas that formed the stars and led to the supernova explosions, we can determine how much gas was used to form stars and how much gas was used to fuel the black hole. The percentage of gas used to fuel the black hole is at most 1 per cent of the gas used to form stars. If on the other hand the outflows from stars fuel the black hole, we conclude that with this supernova rate, the winds from stars would be enough to fuel the AGN during the age of the starburst $\sim 6.5\times10^{7}$ yr. These arguments are of course dependent on how efficient the process of transferring gas from larger scales to the black hole is, but can give us a general overview of the mass budget in the vicinity of the nucleus.

Using the measured kinematics at $r<0{''}.2$, we are able to determine an independent upper limit for the black hole mass of $1.5\times10^{8}$M$_{\odot}$, which is consistent with other estimates. We also reproduce our observations using a dynamical model, and determine an upper limit based on the model predictions for the $V_{\rm rms} = \sqrt{V^{2}+\sigma^{2}}$ in the central $r<0{''}.4$ of the galaxy, of $M_{\rm BH} < 6\times 10^{7}$M$_{\odot}$.

The study of MCG--6-30-15 allowed us to determine the dynamical properties of the inner regions of this galaxy and infer its stellar history. There are a growing number of galaxies that have been observed in detail using integral field spectroscopy. MCG--6-30-15 has a larger mass accretion rate (high AGN activity), which is not common among the usually selected targets due to the difficulty in removing the AGN contamination. We have shown here that it is possible to obtain the stellar kinematics with the presence of Brackett broad emission lines, allowing in the future to increase the parameter range in AGN activity of the galaxies studied.
The kinematically distinct core in the centre of the galaxy may be associated with bar-driven gas inflow, which could, on a smaller scale, be related with the fuelling necessary for AGN activity. We relate the available gas to form stars with the observed AGN activity and obtain general constraints on the mass budget in the centre of the galaxy. Combined studies of stellar properties and AGN activity in the centre of galaxies, will in the future help clarifying the relation between star formation and black hole fuelling at small scales.
\section{Acknowledgements}
The authors would like to thank Roderick Johnstone for useful discussions and the anonymous referee for useful comments that improved this paper.

Some of the images presented in this paper were based on observations made with the NASA/ESA Hubble Space Telescope, and obtained from the Hubble Legacy Archive, which is a collaboration between the Space Telescope Science Institute (STScI/NASA), the Space Telescope European Coordinating Facility (ST-ECF/ESA) and the Canadian Astronomy Data Centre (CADC/NRC/CSA). 

This research has made use of the NASA/IPAC Extragalactic Database (NED) which is operated by the Jet Propulsion Laboratory, California Institute of Technology, under contract with the National Aeronautics and Space Administration.
\bibliographystyle{thesis}
\bibliography{AGN}
\end{document}